\documentclass[aps,pre,preprint]{revtex4-1}
\usepackage[utf8x]{inputenc}
\usepackage{amsmath}
\usepackage{graphicx}
\usepackage{subfig}
\usepackage{sidecap}
\usepackage{verbatim} 
\usepackage{hyperref}

\begin{document}

\title{Crystals and liquid crystals confined to curved geometries}
\author{Vinzenz Koning and Vincenzo Vitelli}
\affiliation{Instituut-Lorentz, Universiteit Leiden, 2300 RA Leiden, The Netherlands}

\begin{abstract} 
\noindent 
This review introduces the elasticity theory of two-dimensional crystals and nematic
liquid crystals on curved surfaces, the energetics of topological defects (disclinations, dislocations
and pleats) in these ordered phases, and the interaction of defects with the
underlying curvature. This chapter concludes with two cases of
three-dimensional nematic phases confined to spaces with curved
boundaries, namely a torus and a spherical shell.  
\end{abstract}

\maketitle
\tableofcontents 
\newpage


\section{Introduction}
Whether it concerns biological matter such as membranes, DNA and
viruses, or synthesised anisotropic colloidal particles, the
deformations inherent to soft matter almost inevitably call for a
geometric description. Therefore, the use of geometry has always been
essential in our understanding of the physics of soft matter. However,
only recently geometry has turned into an instrument for
the design and engineering of micron scaled materials. Key concepts 
are geometrical frustration and the topological defects that are often
a consequence of this frustration \cite{Sadoc, Defects_and_Geometry, RevModPhys.74.953}. 

Geometrical frustration refers to the impossibility of local order to
propagate throughout a chosen space. This impossibility is of
geometric nature and could for instance be due to the topology of the space. Probably your first and most
familiar encounter with this phenomenon was while playing
(association) football. The mathematically inclined amongst you may
have wandered off during the game 
and wondered: ``Why
does the ball contain hexagonal \textit{and} pentagonal panels?" The
ball cannot merely contain hexagonal panels: a perfect tiling of
hexagons (an example of local order) cannot be achieved on the
spherical surface (the space considered). There exists a constraint on
the number of faces, $F$, edges, $E$, and vertices, $V$. The
constraint is named after Euler and reads \cite{Riemann}
\begin{equation}
\label{eq:euler_poly}
F - E + V = \chi,
\end{equation}
where $\chi$ is the Euler characteristic. The Euler characteristic is a quantity insensitive to continuous
deformations of the surface of the ball such as twisting and bending. We call such
quantities topological. Only if
one would perform violent operations such as cutting a hole in the sphere
and glueing a handle to the hole a surface of differently topology
can be created \cite{Riemann, Needham}. For a surface with one handle $\chi=0$, just as for a torus or a
coffee mug. The Euler characteristic $\chi$
equals $2$ for the spherical surface of the ball. Thus, Euler's
polyhedral formula (eq. \eqref{eq:euler_poly}) ensures the need of 12 pentagonal
patches besides the hexagonal ones, no matter how well inflated the
ball is. To see this, write the number of faces $F$ as
the sum of the number of hexagons, $H$, and pentagons, $P$, i.e. $F =
H + P$.  One edge is shared by two faces, hence $E = \frac{1}{2}
\left( 6H - 5P \right)$. Moreover, each vertex is shared among three faces,
hence $V=\frac{1}{3} \left( 6H - 5P \right)$. Substituting the
expressions for $F$, $E$ and $V$ into eq. \eqref{eq:euler_poly} yields
$P=12$. These pentagons are
the defects. Similarly, protein shells of spherical viruses which
enclose the genetic material consist of
pentavalent and hexavalent subunits \cite{virus, PhysRevE.68.051910}. Another condensed matter analog
of the geometrical frustration in footballs is the
`colloidosome'. Colloidosomes are spherical colloidal crystals \cite{2002Sci...298.1006D,2003Sci...299.1716B,2010Natur.468..947I} that are
of considerable interest as microcapsules for delivery and controlled
release of drugs \cite{2002Sci...298.1006D}.

\begin{figure}[h]
\centering
\subfloat{\label{fig:gull}\includegraphics[width=0.3\textwidth]{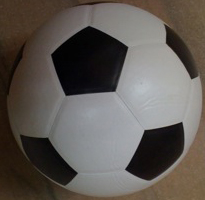}}
\hspace{0.5cm}                
\subfloat{\label{fig:mouse}\includegraphics[width=0.3\textwidth]{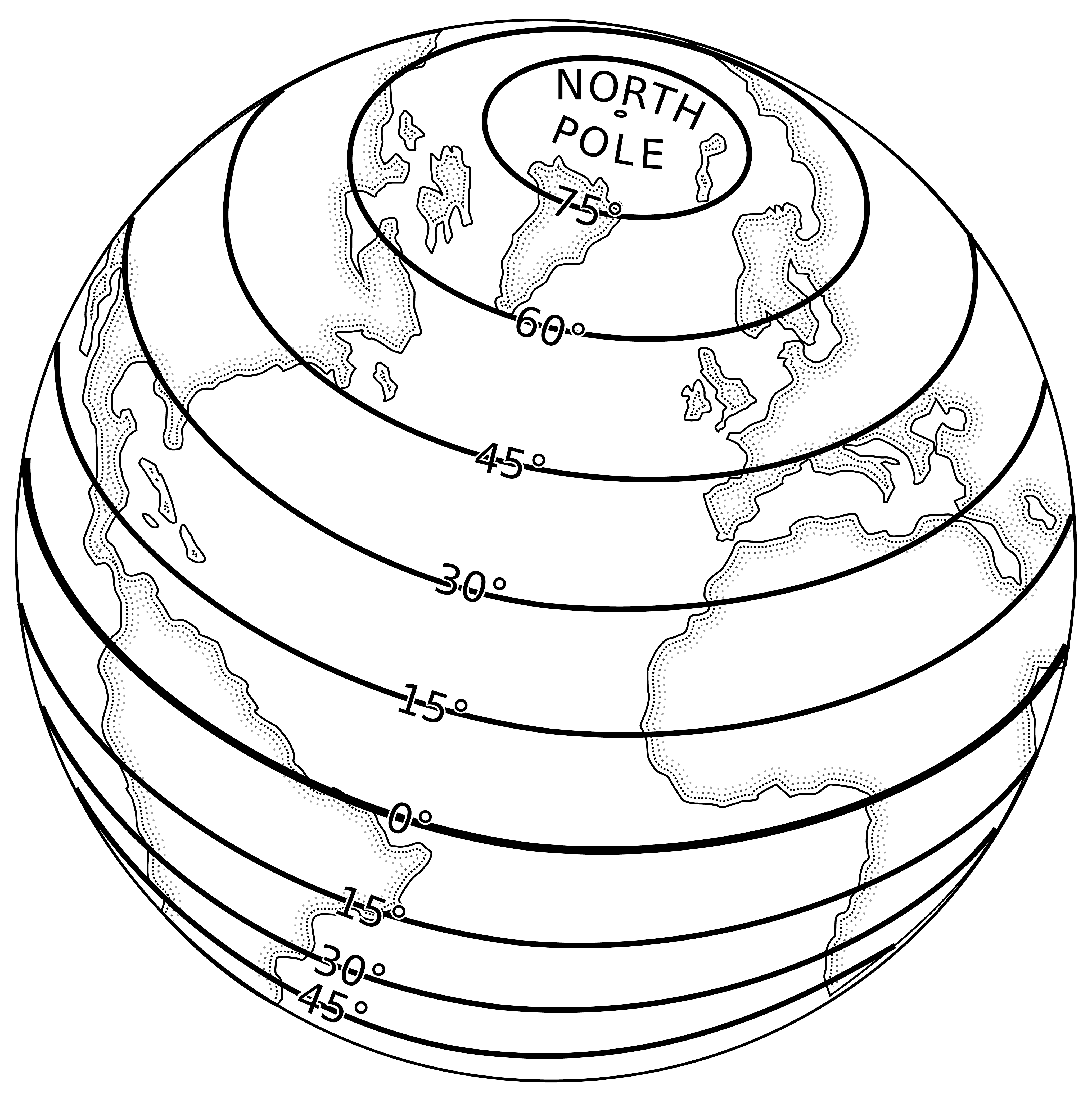}}
\caption{\textit{Left panel:} Geometric frustration in a football. A
  perfect tiling of hexagonal panels cannot be achieved everywhere, resulting in black pentagonal panels (defects). \textit{Right
    panel:} Geometric frustration on the globe. The lines of latitude
  shrink to a point at the north and south poles (defects). Adapted
  from \url{http://commons.wikimedia.org/wiki/File:Latitude_lines}.}
\label{fig:geometricfrustration}
\end{figure}
Another macroscopic example of geometrical frustration are the lines
of latitude on the surface of a globe. The points where these lines
shrink to a point, that is the North and South Poles, are the
defects. Just like the pentagons on the football the defects on the
globe are also required by a topological constraint, namely the
Poincare-Hopf theorem \cite{Needham}: 
\begin{equation}
\sum_a s_a = \chi.
\end{equation}
The lines of latitude circle once around both poles. Hence, there are
two defects with an unit winding number, $s$. (See section \ref{sec:topological_defects} for a
more precise definition.)
Similar to the lines of longitude and latitude on the globe, a coating
of a nanoparticle with a monolayer of ordered tilted molecules also
has two polar defects \cite{PhysRevLett.67.1169,1992JPhy2...2..371L, 2002NanoL...2.1125N, 2006PhRvE..74b1711V, 2007Sci...315..358D}. Recently, Stellacci and co-workers have been
able to functionalise the defects to assemble linear chains of
nanoparticles \cite{2007Sci...315..358D}. A nematic liquid crystal coating possesses four defects
at the vertices of a regular tetrahedron in the ground
state \cite{1992JPhy2...2..371L}. Attaching chemical linkers to these defects could result in a
three-dimensional diamond structure \cite{2002NanoL...2.1125N}, rather than a one-dimensional
chain. This defect arrangement has been recently observed in nematic
double emulsion droplets \cite{2011NatPh...7..391L}, in which a nematic droplet itself contains
another smaller water droplet. However, this is only one of the many
defect arrangements that are observed, as the size and location of the
inner water droplet is varied \cite{2011NatPh...7..391L,C3SM27671F}. Functionalisation of the defects, thus resulting ordered structures confined to curved surfaces or shells offers an intriguing route to directed assembly.      

The types of order that we will discuss in this chapter are
crystalline and (nematic) liquid crystalline. After introducing
mathematical preliminaries, we will discuss the elasticity of crystals
and liquid crystals and give a classification of the defects in these
phases of matter. We will elucidate the role of geometry in this
subject. In particular, we will explicitly show that, in contrast to
the two examples given in the introduction, a topological constraint
is not necessary for geometrical frustration. After that we will explore the fascinating
coupling between defects and curvature. We will briefly comment on the
screening by recently observed
charge-neutral pleats in curved colloidal crystals. We will then cross over from a
two dimensional surface to curved films with a finite thickness and
variations in this thickness. The particular system we are considering is a
spherical nematic shell encapsulated by a nematic double emulsion droplet. We will finish this chapter with a discussion on
nematic droplets of toroidal shape. Though topology does not prescribe
any defects, there is frustration due to the geometric confinement.


\section{Crystalline solids and liquid crystals}
Besides the familiar solid, liquid and gas phases, there exist other
fascinating forms of matter, which display phenomena of order
intermediate between conventional isotropic fluids and crystalline solids. These are therefore called liquid crystalline or mesomorphic
phases \cite{deGennes, RevModPhys.46.617}. Let us consider the
difference between a solid crystal and a liquid crystal. In a
solid crystal all the constituents are located in a periodic fashion,
such that only specific translations return the same lattice. Moreover, the bonds connecting
neighbouring crystal sites define a discrete set of vectors which are
the same throughout the system. In a crystal, there is thus both bond-orientational and translational
order. In liquid crystals there is orientational order, as the anisotropic constituent
molecules define a direction in space, but
the translational order is partially or fully lost. The latter phase,
in which there is no translational order whatsoever, is
called a nematic liquid crystal. The loss of translational order is
responsible for the fluidic properties of nematic liquid crystals. A
thorough introduction to liquid crystals can be found in the chapter
by Lagerwall.


\section{Differential geometry of surfaces}
\subsection{Preliminaries}
For a thorough introduction to the differential geometry of surfaces,
please consult refs. \cite{Struik, David, RevModPhys.74.953}. In this section we will introduce the topic briefly and establish the
notation. Points on a curved surface embedded in the three dimensional world we live in can be described by a three-component vector $\mathbf{R} \left( \mathbf{x} \right)$ as a function of the coordinates $\mathbf{x} = \left( x^1, x^2 \right)$. Vectors tangent to this surface are given by 
\begin{equation}
\label{eq:basis}
\mathbf{t}_{\alpha} = \partial_{\alpha} \mathbf{R},
\end{equation}
where $\partial_{\alpha} = \frac{\partial}{\partial x^{\alpha}}$ is the partial derivative with respect to $x^{\alpha}$. These are in in general neither normalised nor orthogonal. However, it does provide a basis to express an arbitrary tangent vector $\mathbf{n}$ in:
\begin{equation}
\mathbf{n} = n^{\alpha} \mathbf{t}_{\alpha}.
\end{equation} 
Here we have used the Einstein summation convention, \textit{i.e.},
an index occurring twice in a single term is summed over, provided
that one of the them is a lower (covariant) index and the other is an
upper
(contravariant) index. We reserve Greek characters $\alpha$, $\beta$,
$\gamma,\dotso$ as indices for components of vectors and tensors tangent
to the surface. The so-called metric tensor reads
\begin{equation}
g_{\alpha \beta} = \mathbf{t}_{\alpha} \cdot \mathbf{t}_{\beta}.
\end{equation}
and its inverse is defined by
\begin{equation}
g^{\alpha \beta} g_{\beta \gamma} = \delta^{\alpha}_{\gamma},
\end{equation}
where $\delta^{\alpha}_{\gamma}$ is equal to one if $\alpha = \gamma$
and zero otherwise. We can lower and raise indices with the
metric tensor and inverse metric tensor, respectively, in the usual way, e.g.
\begin{equation}
g_{\alpha \beta} n^{\alpha} = n_{\beta}
\end{equation} 
It is straightforward to see that the inner product between two vectors $\mathbf{n}$ and $\mathbf{m}$ is 
\begin{equation}
\mathbf{n} \cdot \mathbf{m} = n^{\alpha} \mathbf{t}_\alpha \cdot m^{\beta} \mathbf{t}_\beta = g_{\alpha \beta} n^{\alpha} m^{\beta} = n^{\alpha} m_{\alpha}. 
\end{equation}
The area of the parallelogram generated by the infinitesimal vectors $dx^1 \mathbf{t}_1$ and $dx^2 \mathbf{t}_2$, given by the magnitude of their cross product, yields the area element 
\begin{align}
dS &= \left| dx^1 \mathbf{t}_1 \times dx^2 \mathbf{t}_2 \right|  \nonumber\\ &= \sqrt{ \left( \mathbf{t}_1 \times \mathbf{t}_2 \right)^2} dx^1 dx^2    
\nonumber\\ &= \sqrt{ \left| \mathbf{t}_1 \right|^2 \left| \mathbf{t}_2 \right|^2 - \left( \mathbf{t}_1 \cdot \mathbf{t}_2 \right)^2} dx^1 dx^2 
\nonumber\\ &= \sqrt{ g_{11} g_{22} - g_{12} g_{21} }dx^1 dx^2
\nonumber\\ &= \sqrt{ g } d^2 x 
\end{align}
where $g = \mbox{det}(g_{\alpha \beta})$, the determinant of the metric tensor, and $d^2 x$ is shorthand for $dx^1 dx^2$.
More generally, the magnitude of the cross product of two vectors $\mathbf{m}$ and $\mathbf{n}$ is 
\begin{equation}
\left| \mathbf{m} \times \mathbf{n} \right| = \left| \gamma_{\alpha \beta} m^{\alpha} n^{\beta} \right|,
\end{equation}
which introduces the antisymmetric tensor 
\begin{equation}
\gamma_{\alpha \beta} = \sqrt{g} \epsilon_{\alpha \beta}
\end{equation}
where $\epsilon_{\alpha\beta}$ is the Levi-Civita symbol satisfying $\epsilon_{1 2} = -\epsilon_{2 1} = 1$ and is zero otherwise.

Since we will encounter tangent unit vectors, e.g. indicating the orientation of some physical quantity, it is convenient to decompose this vector in a set of  orthonormal tangent vectors, $\mathbf{e}_1 \left( \mathbf{x} \right)$ and $\mathbf{e}_2 \left( \mathbf{x} \right)$, such that 
\begin{equation}
\mathbf{e}_i \cdot \mathbf{e}_j = \delta_{ij} \quad \text{and} \quad \mathbf{N} \cdot \mathbf{e}_i = 0,
\end{equation}
alternative to the basis defined in eq. \ref{eq:basis}. Here
$\mathbf{N}$ is the vector normal to the surface. We use the 
Latin letters $i$, $j$ and $k$ for the components of vectors expressed
in this orthonormal basis. As they are locally Cartesian they do not
require any administration of the position of the index.  
Besides the area element we need a generalisation of the partial
derivative. This generalisation is the covariant derivative,
$D_\alpha$, the projection of the derivative onto the surface. The
covariant derivative of $\mathbf{n}$ expressed in the orthonormal
basis reads in component form \cite{RevModPhys.74.953}
\begin{align}
D_\alpha n_i 
& = \mathbf{e}_i \cdot \partial_\alpha \mathbf{n} \nonumber \\ 
& = \mathbf{e}_i \cdot \partial_\alpha n_j \mathbf{e}_j + \mathbf{e}_i \cdot \partial_\alpha \mathbf{e}_j n_j \nonumber\\
& = \partial_\alpha n_i + \epsilon_{ij} A_\alpha n_j,   
\end{align}
where $\epsilon_{ij} A_\alpha = \mathbf{e}_i \cdot \partial_\alpha
\mathbf{e}_j$ is called the spin-connection. The final line is
justified because the derivative of any unit vector is perpendicular to this unit vector. 
More generally, the covariant derivative of the vector $\mathbf{n}$ along $x^\alpha$ is \cite{David}
\begin{equation}
D_{\alpha}n^\beta = \partial_{\alpha} n^\beta + \Gamma^\beta_{\alpha \gamma} n^{\gamma}
\end{equation}
where the Christoffel symbols are 
\begin{equation}
\Gamma^\alpha_{\beta \gamma} = \frac{1}{2} g^{\alpha \delta} \left( \partial_\gamma g_{\beta \delta} + \partial_\beta g_{\delta \gamma} - \partial_\delta g_{\beta \gamma} \right).
\end{equation}
Finally, with the antisymmetric tensor and the area element in hand we can state a useful formula in integral calculus, namely Stokes' theorem
\begin{equation}
\int d^2 x \sqrt{ g } \gamma^{\alpha \beta} D_{\alpha} n_{\beta} = \oint dx^{\alpha} n_{\alpha}. 
\end{equation}

\subsection{Curvature}
\begin{figure}[h]
\centering
\includegraphics[width=0.4\textwidth]{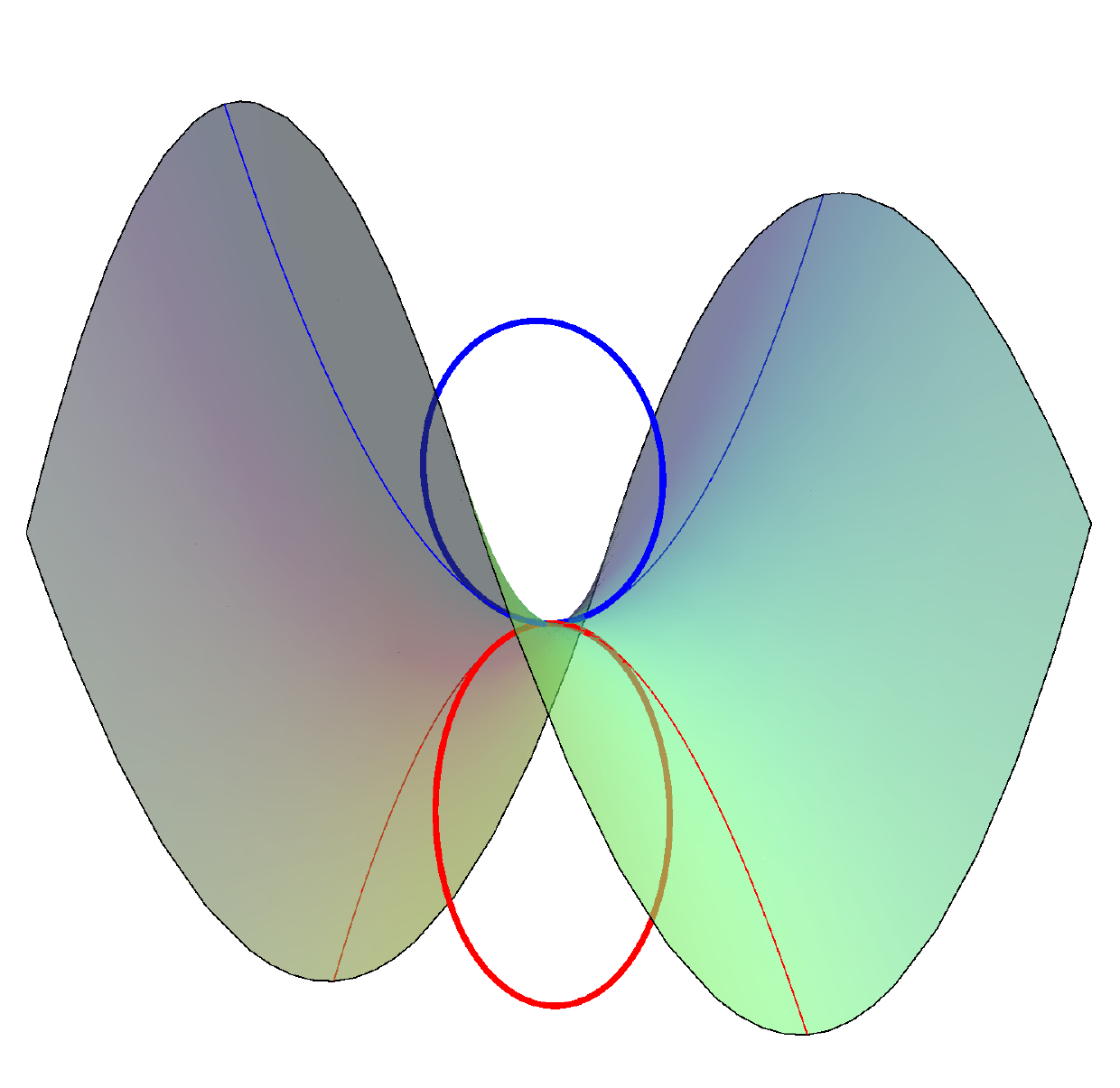}
\caption{Saddle surface has negative Gaussian curvature. $\kappa_1 $
  and $ \kappa_2 $ have different signs. Tangent circles are drawn in
  blue and red.}
\label{fig:saddle_drawing}
\end{figure}

The curvature is the deviation from flatness and therefore a measure
of the rate of change of the tangent vectors along the normal, or, put
the other way around, a measure of the rate of change of the normal
along the tangent vectors. This can be cast in a curvature tensor defined as
\begin{equation}
\label{eq:curvature_tensor}
K_{\alpha \beta} = \mathbf{N} \cdot \partial_{\beta}
\mathbf{t}_\alpha.  = -\mathbf{t}_\alpha \cdot \partial_{\alpha} \mathbf{N} 
\end{equation}
From this tensor we extract the intrinsic Gaussian curvature
\begin{equation}
\label{eq:gaussian}
G = \mbox{det} \left( K^{\alpha}_{\beta} \right) = \frac{1}{2} \gamma^{\alpha \beta} \gamma^{\gamma \delta} K_{\alpha \beta} K_{\gamma \delta} = \kappa_1 \kappa_2
\end{equation}
and extrinsic mean curvature
\begin{equation}
H = \frac{1}{2} \mbox{Tr} \left(K^{\alpha}_{\beta} \right) = \frac{1}{2} g^{\alpha \beta} K_{\alpha \beta} = \frac{1}{2} \left( \kappa_1 + \kappa_2 \right),
\end{equation} 
where $\kappa_1 = \mathbf{N} \cdot \partial_1 \mathbf{\tilde{e}}_1$
and $\kappa_2 = \mathbf{N} \cdot \partial_2 \mathbf{\tilde{e}}_2$ are
the extremal or principal curvatures, the curvature in the principal
directions $\mathbf{\tilde{e}}_1$ and $\mathbf{\tilde{e}}_2$. These
eigenvalues and eigenvectors can be obtained by diagonalising the
matrix associated with the curvature tensor.  If at a point on a
surface $\kappa_1 $ and $ \kappa_2 $ have the same sign the Gaussian
curvature is positive and from the outsiders' point of view the
surface curves away in the same direction whichever way you go, as is
the case on tops and in valleys. In contrast, if at a point on a
surface $\kappa_1 $ and $ \kappa_2 $ have opposite signs the Gaussian
curvature is negative, the  saddle-like surface curves away in
opposite directions. The magnitude of $\kappa_1 $ and $ \kappa_2 $ is
equal to the inverse of the radius of the tangent circle in the
principal direction (Fig. \ref{fig:saddle_drawing}).
It turns out that the Gaussian curvature and the spin-connection are
related. We will see how in a moment by considering the normal (third)
component of the curl (denoted by $\nabla \times$) of the spin-connection
\begin{align}
\left( \nabla \times \mathbf{A} \right)_3 & = \epsilon_{3jk} \partial_j
\left( \mathbf{e}_1 \cdot \partial_k \mathbf{e}_2 \right) \nonumber
\\ & = \epsilon_{3jk} \partial_j \mathbf{e}_1 \cdot \partial_k
\mathbf{e}_2 \nonumber
\\ & = \epsilon_{3jk} \left( \mathbf{N} \cdot \partial_j \mathbf{e}_1 \right) \left( \mathbf{N} \cdot \partial_k \mathbf{e}_2 \right) 
\end{align}
where we have used the product rule and the antisymmetry of
$\epsilon_{ijk}$ in the second equality sign. The final line is
justified by the fact that the derivative of a unit vector is
perpendicular to itself and therefore we have e.g. $\partial_j
\mathbf{e}_1 = \left( \mathbf{N} \cdot \partial_j \mathbf{e}_1 \right)
\mathbf{N} + \left( \mathbf{e}_2 \cdot \partial_j \mathbf{e}_1 \right)
\mathbf{e}_2$. If we now with the aid of eqs. \eqref{eq:gaussian} and
\eqref{eq:curvature_tensor} note that
\begin{equation}
G = \left( \mathbf{N} \cdot \partial_1 \mathbf{e}_1 \right) \left( \mathbf{N} \cdot \partial_2 \mathbf{e}_2 \right)
- \left( \mathbf{N} \cdot \partial_1 \mathbf{e}_2 \right) \left( \mathbf{N} \cdot \partial_2 \mathbf{e}_1 \right)
\end{equation}      
we easily see that the normal component of the curl of the spin-connection equals the Gaussian curvature:
\begin{equation}
\left( \nabla \times \mathbf{A} \right) \cdot \mathbf{N} = G,
\end{equation}
or alternatively\cite{refId0}
\begin{equation}
\label{eq:curl_connection_gauss}
\gamma^{\alpha \beta} D_\alpha A_\beta = G. 
\end{equation}
This geometrical interpretation of $\mathbf{A}$ will show its
importance in section \ref{sec:elasticity_on_curved_surfaces}, where
we will comment on its implications on the geometrical frustration in curved nematic liquid crystal films. 

\subsection{Monge gauge}
A popular choice of parametrisation of the surface is the Monge gauge or height representation in which $\mathbf{x} = \left( x, y \right)$ and $\mathbf{R} = \left( x, y, f \left(x,y \right) \right)$, where $f \left(x,y \right)$ is the height of the surface above the $xy$-plane. In this representation the Gaussian curvature reads
\begin{equation}
\label{eq:gaussian_monge}
G = \frac{\mbox{det } \partial_\alpha \partial_\beta f}{g},
\end{equation}
where the determinant of the metric is given by
\begin{equation}
g = 1 + \left( \partial_x f \right)^2 + \left( \partial_y f \right)^2 .
\end{equation}


\section{Elasticity on curved surfaces and in confined geometries}
\label{sec:elasticity_on_curved_surfaces}
\subsection{Elasticity of a two-dimensional nematic liquid crystal}
\begin{figure}[h]
\centering
\includegraphics[width=0.2\textwidth]{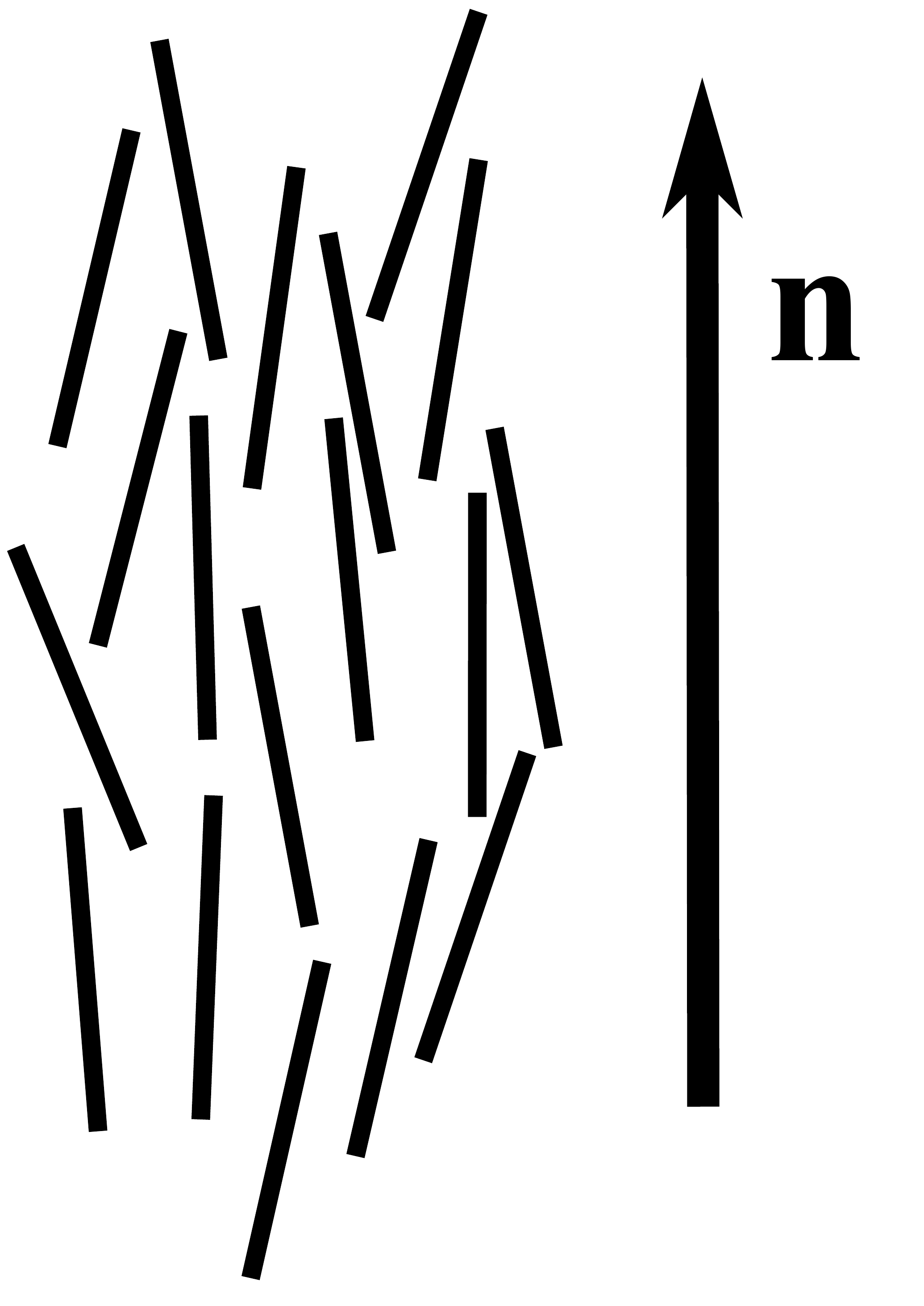}
\caption{The director $\mathbf{n}$ specifies the average local orientation of the nematic molecules.}
\label{fig:nematagons}
\end{figure}

\begin{figure}[h]
\centering
\includegraphics[width=0.4\textwidth]{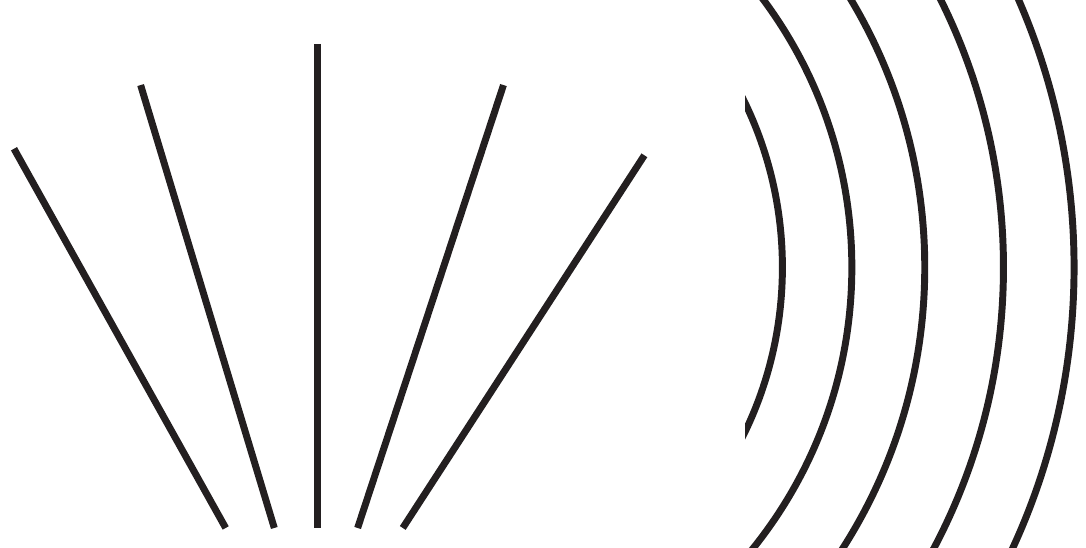}
\caption{Conformations with (left panel) a non-vanishing divergence of
  the director and (right panel) a non-vanishing curl of the director.}
\label{fig:deformations}
\end{figure}

In a nematic liquid crystal the molecules (assumed to be anisotropic) tend to align parallel to a common axis. The direction of this axis is labeled with a unit vector, $\mathbf{n}$, called the director (see Fig. \ref{fig:nematagons}). The states $\mathbf{n}$ and $-\mathbf{n}$ are equivalent.
Any spatial distortion of a uniform director field costs energy. If we assume that these deformations are small on the molecular length scale, $l$, 
\begin{equation}
\lvert \partial_i n_j \rvert \ll \frac{1}{l}, 
\end{equation}
we can construct a phenomenological continuum theory. The resulting Frank free energy $F$ for a two dimensional flat nematic liquid crystal reads \cite{deGennes, Kleman, SoftMatter}
\begin{equation}
\label{eq:Frank_2d_flat}
F = \frac{1}{2} \int  d^2x \left[ k_1 \left( \partial_{i} n_{i} \right)^2   + k_3 \left( \epsilon_{ij} \partial_{i} n_{j}  \right)^2 \right] ,
\end{equation}
where the splay and bend elastic constants, $k_1$ and $k_3$ respectively, measure the energy of the two independent distortions shown in Fig. \ref{fig:deformations}. 
To simplify the equations one often makes the assumption of isotropic elasticity. In this approximation the Frank elastic constants are equal, $k_1 = k_3 = k$, and up to boundary terms the free energy reduces to \cite{Kleman}
\begin{equation}
\label{eq:energy 2d}
F = \frac{1}{2} k \int d^2x  \partial_i n_j \partial_i n_j, 
\end{equation}
When the coupling of the director to the curvature tensor
$K_{\alpha\beta}$ \cite{PhysRevLett.99.017801, PhysRevE.80.051703,PhysRevE.76.031908, PhysRevE.77.041705,
  PhysRevLett.108.017801,doi:10.1021/jp205128g,
  PhysRevLett.108.207803,PhysRevE.85.061701,Napoli201366} is ignored,
the elastic free energy on a curved surface generalises to
\cite{PhysRevLett.67.1169, PhysRevE.53.2648, 2002NanoL...2.1125N,
  PhysRevE.70.051105, 2006PhRvE..74b1711V}
\begin{equation}
F = \frac{1}{2} k \int d^2x \sqrt{g}  D_\alpha n^\beta D^\alpha n_\beta, 
\end{equation} 
In this equation the area element has become $dS = d^2 x \sqrt{g}$ and partial derivatives have been promoted to covariant derivatives.
Because the director is of unit length, we can conveniently specify it in terms of its angle with a local orthonormal reference frame, $\Theta \left( \mathbf{x} \right)$, as follows
\begin{equation}
\mathbf{n} = \cos \left( \Theta \right) \mathbf{e_1}+ \sin \left( \Theta \right) \mathbf{e_2}.
\end{equation}
Then, since $\partial_\alpha n_1 = - \sin \left( \Theta \right) \partial_\alpha \Theta = - n_2 \partial_\alpha \Theta$ and $\partial_\alpha n_2 = \cos \left( \Theta \right) \partial_\alpha \Theta = n_1 \partial_\alpha \Theta$ we see that
\begin{equation}
\partial_\alpha n_i = - \epsilon_{ij} n_j \partial_\alpha \Theta
\end{equation}
with which we find the covariant derivative to be
\begin{equation}
D_\alpha n_i = - \epsilon_{ij} n_j \left( \partial_{\alpha} \Theta - A_{\alpha} \right) 
\end{equation}
Therefore, we can rewrite the elastic energy as\cite{refId0}
\begin{equation}
\label{eq:energy with connection}
F=\frac{1}{2} k \int d^2 x \sqrt{g} \left( \partial_{\alpha} \Theta - A_{\alpha}\right) \left( \partial^{\alpha} \Theta - A^{\alpha}\right), 
\end{equation}
where we have used that $\left( - \epsilon_{ij} n_j \right) \left( -
  \epsilon_{ij} n_j \right) = \delta_{jk} n_j n_k = \cos^2 \left(
  \Theta \right) + \sin^2 \left( \Theta \right) = 1$. This form of the free energy 
\footnote{Note that if we had chosen orthonormal reference frame differing by a local rotation $\Psi \left( \mathbf{x} \right) $ 
\begin{equation}
\mathbf{e}_1 \left( \mathbf{x} \right) \rightarrow \cos \left( \Psi \left( \mathbf{x} \right) \right) \mathbf{e}_1 \left( \mathbf{x} \right) - \sin \left( \Psi \left( \mathbf{x} \right) \right) \mathbf{e}_2 \left( \mathbf{x} \right)
\end{equation}
\begin{equation}
\mathbf{e}_2 \left( \mathbf{x} \right) \rightarrow \sin \left( \Psi \left( \mathbf{x} \right) \right) \mathbf{e}_1 \left( \mathbf{x} \right) + \cos \left( \Psi \left( \mathbf{x} \right) \right) \mathbf{e}_2 \left( \mathbf{x} \right)
\end{equation}
implying 
\begin{equation}
\Theta \left( \mathbf{x} \right) \rightarrow \Theta \left( \mathbf{x} \right) + \Psi \left( \mathbf{x} \right) \quad A_\alpha \left( \mathbf{x} \right) \rightarrow A_\alpha \left( \mathbf{x} \right) + \partial_\alpha \Psi \left( \mathbf{x} \right) 
\end{equation} 
the free energy, eq. \eqref{eq:energy with connection}, remains the same.}
clearly shows that nematic order on
curved surface is geometrically frustrated. The topological
constraints of the introductory section are merely a special example
of the frustration of local order due to the geometrical properties of
the system. Note that for a curved surface without such a topological
constraint (\textit{e.g.} a Gaussian bump) the
ground state can be a deformed director field. Since the curl of the
spin-connection equals the Gaussian curvature (eq. \eqref{eq:curl_connection_gauss}), 
if the gaussian curvature is nonzero, the spin-connection is
irrotational and cannot be written as the gradient of a scalar field, $A_{\alpha} \neq \partial_{\alpha} \Theta$,
just like the magnetic field cannot be described by a scalar field either.
Therefore $F$ in eq. \ref{eq:energy with connection} is nonzero and we can conclude that there is geometrical
frustration present in the system.

\subsection{Elasticity of a two-dimensional solid}
Similar to the construction of the continuum elastic energy of a
nematic liquid crystal, we can write down the elastic energy of a
linear elastic solid as an integral of terms quadratic in the
deformations, \textit{i.e.} strain. This strain is found in the
following way. Consider a point $\mathbf{x}=\left(x,y,0\right)$ on an
initially flat solid. This point is displaced to
$\mathbf{x}' \left( \mathbf{x} \right)=\left(x',y',f\right)$ in the deformed
solid, and so we may define a
displacement vector
$\mathbf{u}\left(\mathbf{x}\right)=\mathbf{x}'-\mathbf{x}=u_x
\mathbf{e}_x + u_y \mathbf{e}_y + f \mathbf{e}_z$. The square
of the line element in the deformed plate is then given by $ds'^2 =
\left( dx + du_x\right)^2 + \left( dx + du_x\right)^2 + df^2$. Noting
that $du_x = \partial_i u_x dx_i$ with $x_i=x,y$ and similarly for $u_y$
and $f$ we find \cite{Landau}
\begin{equation}  
ds'^2 = ds^2 + 2 u_{ij} dx_i dx_j.
\end{equation}
Thus, the strain tensor $ u_{ij} \left( \mathbf{x}\right)$ encodes how infinitesimal distances
change in the deformed body with respect to the resting state of the
solid and reads
\begin{equation}
\label{eq:strain}
u_{ij}=\frac{1}{2} \left( \partial_i u_j + \partial_j u_i + A_{ij}\right),
\end{equation}
where we have omitted non-linear terms of second order in $\partial_i
u_j$ and where the tensor field $A_{ij}\left(\mathbf{x}\right)$ is now defined as 
\begin{equation}
A_{ij} \equiv \partial_i f \partial_j f.
\end{equation}
We will assume that curvature plays its part only through this coupling of gradients of the displacement field to
the geometry of the surface, and we will therefore adopt the flat
space metric. This is a valid approximation for moderately curved
solids, as we comment on at
the end of the section \cite{Vitelli15082006,2009AdPhy..58..449B}. To leading order in gradients of the height
function, $A_{ij}$ is related to the curvature as (see eq. \eqref{eq:gaussian_monge})
\begin{equation}
\label{eq:A_and_curvature}
- \frac{1}{2} \epsilon_{ik} \epsilon_{jl} \partial_k \partial_l A_{ij} = \mbox{det} \left( \partial_i \partial_j f \right) = G.   
\end{equation} 
Isotropy of the solid leaves two independent scalar combinations of $u_{ij}$
that contribute
to the stretching energy: \cite{Landau}
\begin{equation}
F = \frac{1}{2}\int dS \left( 2\mu u_{ij}^2 + \lambda u_{ii}^2 
\right).
\end{equation}
The elastic constants $\lambda$ and $\mu$ called the Lame
coefficients. 
Minimisation of this energy with respect to $u_j$ leads to the force balance equation:
\begin{equation}
\partial_i \sigma_{ij} = 0,
\end{equation}
where the stress tensor $\sigma_{ij} \left( \mathbf{x} \right)$ is defined by Hooke's law 
\begin{equation}
\label{eq:Hooke}
\sigma_{ij} = 2 \mu u_{ij} + \lambda \delta_{ij} u_{kk}.
\end{equation}
The force balance equation can be solved by introducing the Airy stress function, $\chi \left( \mathbf{x} \right)$, which satisfies
\begin{equation}
\sigma_{ij}  = \epsilon_{ik} \epsilon_{jl} \partial_k \partial_l \chi,
\end{equation}
since this automatically gives
\begin{equation}
\label{eq:force_balance_solved}
\partial_i \sigma_{ij} = \epsilon_{jk} \partial_k \left[ \partial_1, \partial_2 \right] \chi= 0
\end{equation}
by the commutation of the partial derivatives. If one does not adopt
the flat space metric, the covariant generalisation of the force balance equation is not satisfied,
because the the commutator of the covariant derivatives, known as the
Riemann curvature tensor, does not vanish. It is actually proportional
to the Gaussian curvature and indicates why the range of validity of
this approach is limited to moderately curved surfaces \cite{Vitelli15082006,2009AdPhy..58..449B}.
Finally, for small $\partial_i u_j$ 
the bond angle field, $\Theta \left( \mathbf{x} \right)$,
is given by
\begin{equation}
\label{eq:bondangle}
\Theta = \frac{1}{2} \epsilon_{ij} \partial_i u_j.
\end{equation}

\subsection{Elasticity of a three-dimensional nematic liquid crystal}
Besides splay and bend, there are two other deformations possible in a
three dimensional nematic liquid crystal. They are twist and
saddle-splay, measured by elastic moduli $K_2$ and $K_{24}$. The
analog of eq. \eqref{eq:Frank_2d_flat} reads
\begin{equation}
\label{eq:Frank_in_3D}
\begin{split}
F[\mathbf{n}\left(\mathbf{x}\right)] =& \frac{1}{2} \int dV \left( K_1 \left( \nabla \cdot \bf{n} \right)^2
+ K_2 \left( \mathbf{n} \cdot \nabla \times \mathbf{n} \right)^2 \right.  \\
&+ \left. K_3 \left( \mathbf{n} \times \nabla \times \mathbf{n} \right)^2 \right)
- K_{24} \int \mathbf{dS} \cdot \left( \mathbf{n} \nabla \cdot
  \mathbf{n} +   \mathbf{n} \times \nabla \times \mathbf{n} \right).
  \end{split}
\end{equation}
The integration of the splay, twist and bend energy density is over the volume to which the nematic is confined. The
saddle-splay energy per unit volume is a pure divergence term, hence
the saddle-splay energy can be written as the surface integral in
eq. \eqref{eq:Frank_in_3D}. In addition to the energy in eq. \eqref{eq:Frank_in_3D},
there is an energetic contribution coming from the interfacial interactions, often
larger in magnitude. Therefore, the anchoring of the nematic molecules
at the boundary can be taken as a constraint. In one of the possible
anchoring conditions the director is forced to be tangential to the surface, yet
free to rotate in the plane. In this case, the saddle-splay energy
reduces to \cite{2013arXiv1312.5092K}
\begin{equation}
F_{24} = K_{24}  \int dS \left( \kappa_1  n_1^2 + \kappa_2  n_2^2 \right),
\end{equation}
thus coupling the director to the boundary surface.
We refer to the chapter by Lagerwall for a
more detailed discussion on the origin of eq. \eqref{eq:Frank_in_3D}.


\section{Topological defects}
\label{sec:topological_defects}
Topological defects are characterised by a small region where the
order is nod defined. Topological defects in
translationally ordered media, such as crystals, are called
\textit{dislocations}. Defects in the orientational order,
such as in nematic liquid crystals and again crystals, are called
\textit{disclinations}. The defects are topological when they cannot
be removed by a continuous deformation of the order parameter. As we
will see momentarily, they are classified according to a topological
quantum number or topological charge, a quantity that may only take on
a discrete set of values and which can be measured on any circuit
surrounding the defect.

\subsection{Disclinations in a nematic}
\begin{figure}[h]
\centering
\subfloat[$s=1$, $c=0$]{\label{fig:a}\includegraphics[width=0.25\textwidth]{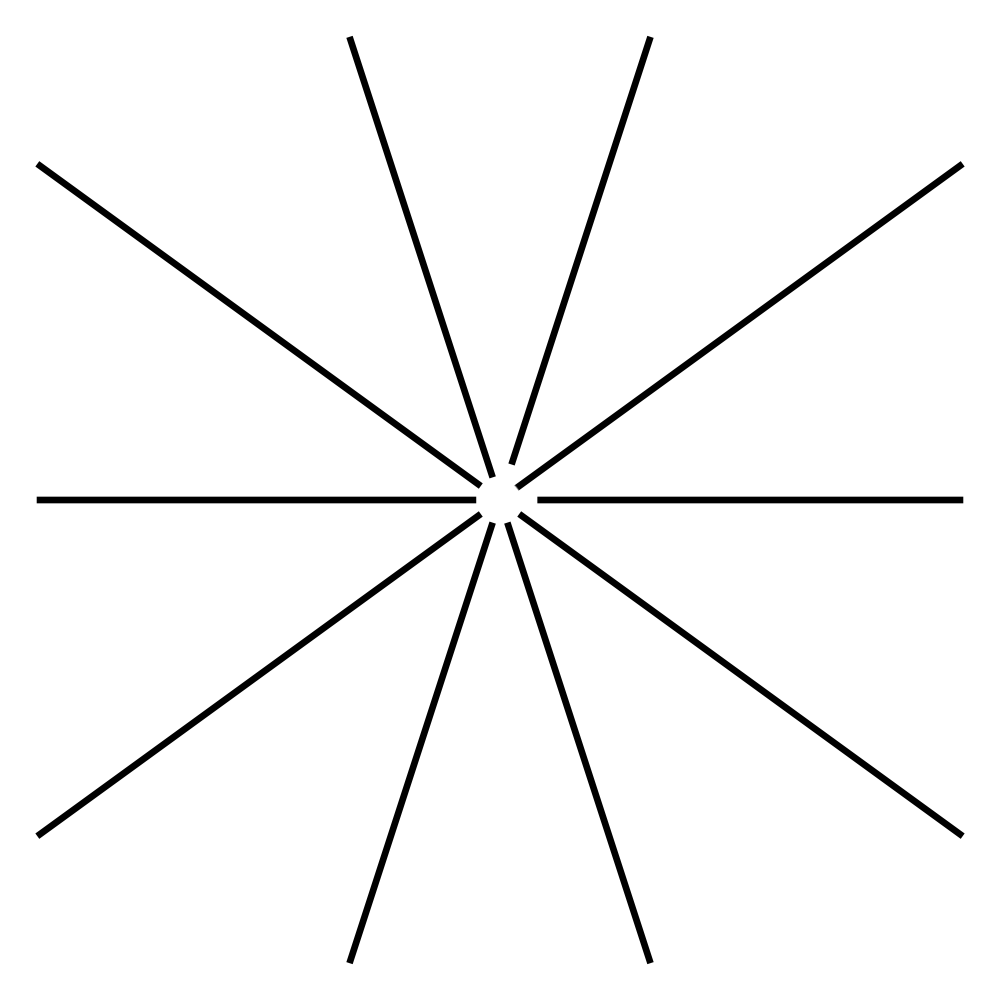}}                
\subfloat[$s=1$, $c=\frac{\pi}{4}$]{\label{fig:b}\includegraphics[width=0.25\textwidth]{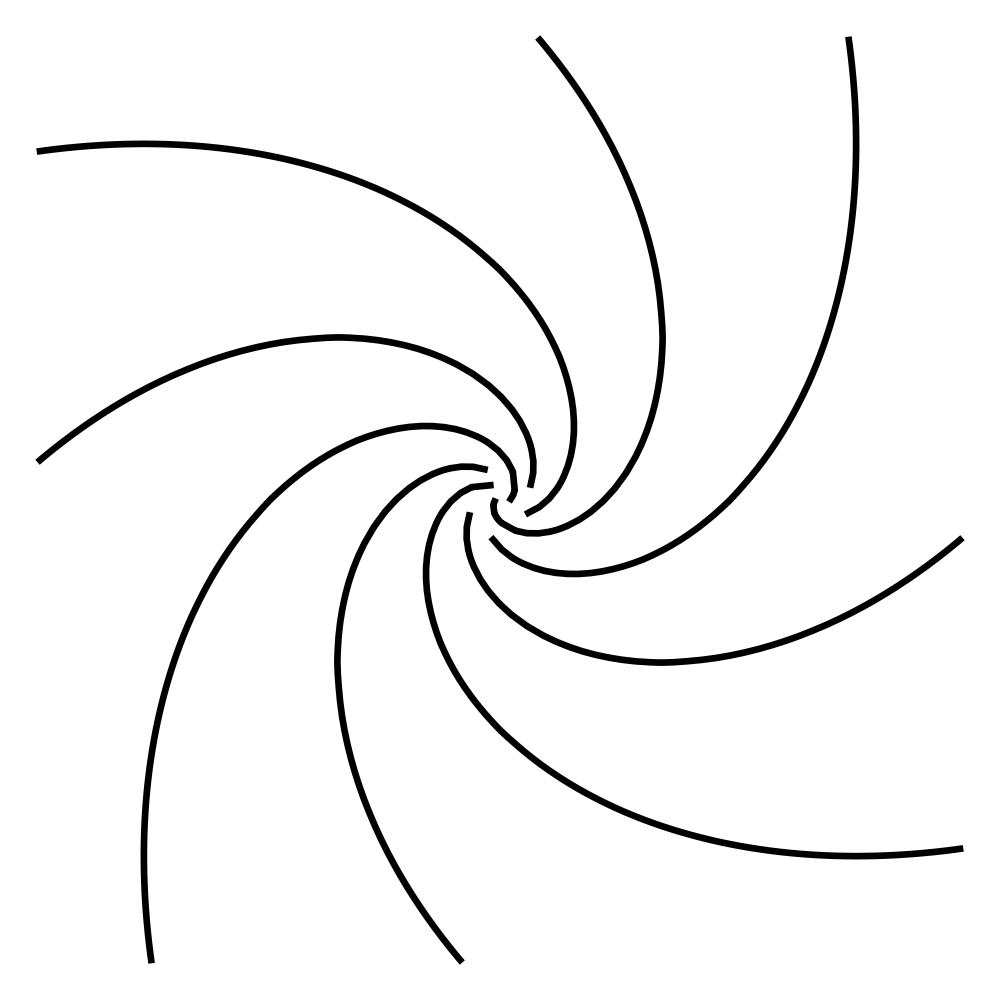}}
\subfloat[$s=1$, $c=\frac{\pi}{2}$]{\label{fig:c}\includegraphics[width=0.25\textwidth]{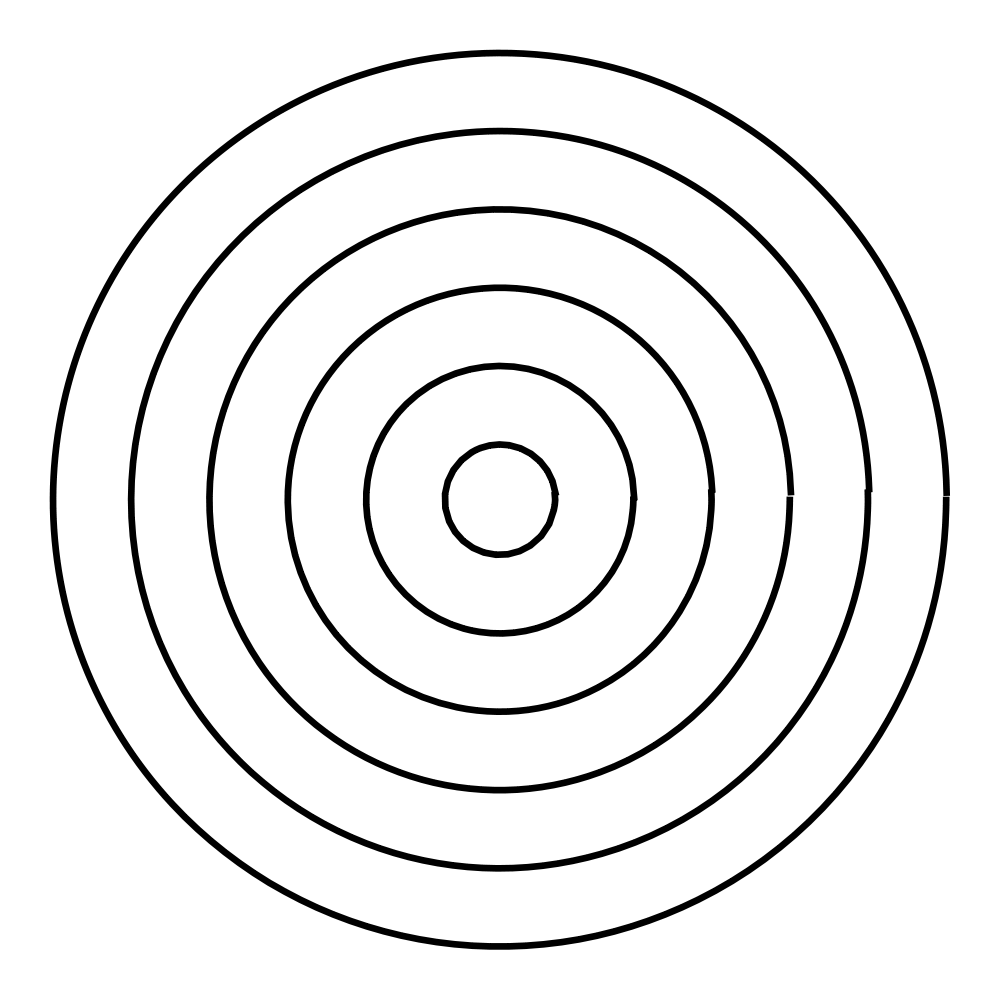}}
\subfloat[$s=\frac{1}{2}$, $c=0$]{\label{fig:d}\includegraphics[width=0.25\textwidth]{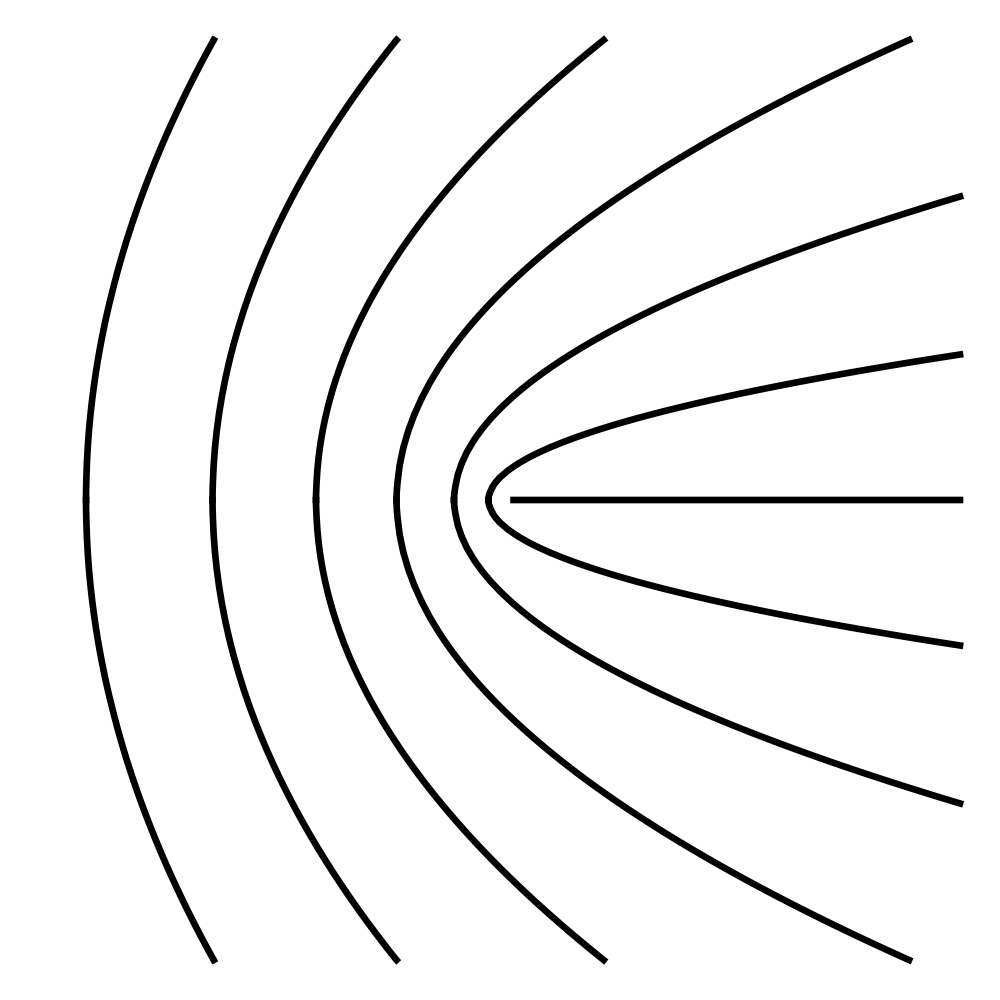}}
\caption{Director configurations, $n_1=\cos \Phi$, $n_2=\sin \Phi$, for disclinations of strength $s$ and constant $c$.}
\label{fig:defects}
\end{figure}

Consider for concreteness a two dimensional nematic liquid crystal. A singularity in the director field is an example of a disclination. Such a point defect can be classified by its winding number, strength, or topological charge, $s$, which is the number of times the director rotates by $2\pi$, when following one closed loop in counterclockwise direction around the singularity: 
\begin{equation}
\label{eq:disclination_nematic}
\oint d \Theta = \oint dx^\alpha \partial_\alpha \Theta = 2 \pi s
\end{equation}
We can express eq. \eqref{eq:disclination_nematic} in differential
form by invoking Stokes' theorem: 
\begin{equation}
\label{eq:disclination_nematic_diff}
\gamma^{\alpha \beta} D_\alpha \partial_\beta \Theta= q \delta \left( \mathbf{x} - \mathbf{x}_a\right)  
\end{equation}
where we use an alternative labelling, $q=2 \pi s$, of the charge of the defect, which is located at $\mathbf{x}_a$. The delta-function obeys
\begin{equation}
\delta \left( \mathbf{x} - \mathbf{x}_a\right) = \frac{\delta \left( x^1 - x_a^1\right) \delta \left( x^2 - x_a^2\right)}{\sqrt{g}},
\end{equation}
such that the integral over the surface yields one.
The far field contribution of the defect to the angular director in a flat plane reads
\begin{equation}
\Theta = s \phi + c,
\label{eq:director point defect}
\end{equation}
as it forms a solution to the Euler-Lagrange equation of the elastic free energy 
\begin{equation}
\partial^2 \Theta = 0.
\label{eq:laplace}
\end{equation}
Here, $\phi$ is the azimuthal angle and $c$ is just a phase. 
Examples are presented in Fig. \ref{fig:defects}. 
Note that since the states $\mathbf{n}$ and $-\mathbf{n}$ are equivalent, defects with half-integer strength are also possible. In fact, it is energetically favourable for an $s=1$ defect to unbind into two $s=\frac{1}{2}$ defects \cite{2002NanoL...2.1125N, Principles}. 

\subsection{Disclinations in a crystal}
\begin{figure}[h]
\centering
\includegraphics[width=0.7\textwidth]{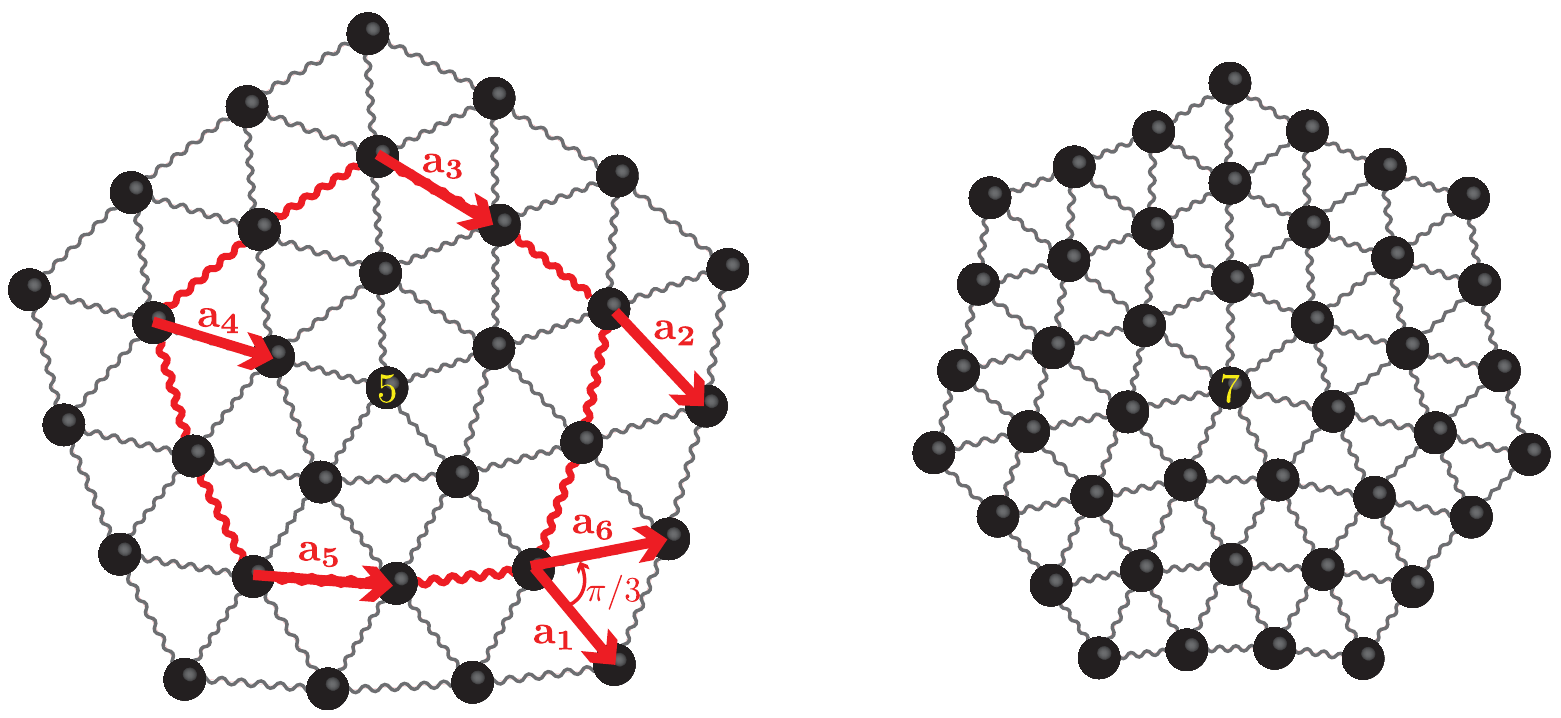}
\caption{(Left panel) Five-fold and (right panel) seven-fold
  disclination. When following a closed counterclockwise loop (red) around
  the five-fold disclination, the initial lattice vector
  $\mathbf{a}_1$ rotates via $\mathbf{a}_2$, $\mathbf{a}_3$,
  $\mathbf{a}_4$ and $\mathbf{a}_5$ over an
  angle of $\pi / 3$ to $\mathbf{a}_6$.}
\label{fig:five_seven_disclination}
\end{figure}

Though energetically more costly, disclinations also arise in
two-dimensional crystals. At these points the coordination number
deviates from its ordinary value, which is six for a crystal on a
triangular lattice. Just like in nematic liquid crystals,
disclinations in crystals are labelled by a topological charge, $q$,
which is the angle over which the vectors specifying the lattice
directions rotate when following a counterclockwise circuit around the
disclination. If we parametrise these lattice direction vectors with
$\Theta \left( \mathbf{x} \right)$, the bond-angle field, this condition reads mathematically
\begin{equation}
\label{eq:disclination_crystal}
\oint d \Theta = q.
\end{equation}
Thus for disclinations in a triangular lattice with five-fold and
seven-fold symmetry, as displayed in
Fig. \ref{fig:five_seven_disclination}, $q=\frac{\pi}{3}$ and $q=-\frac{\pi}{3}$ respectively. 
Analogous to eq. \eqref{eq:disclination_nematic_diff}, the flat-space differential form of
eq. \eqref{eq:disclination_crystal} for a disclination located at $\mathbf{x}_a$ reads
\begin{equation}
\label{eq:disclination_crystal_diff}
\epsilon_{ij} \partial_i \partial_j \Theta = q \delta \left( \mathbf{x} - \mathbf{x}_a\right) 
\end{equation}

\subsection{Dislocations}
\begin{figure}[h]
\centering
\includegraphics[width=0.4\textwidth]{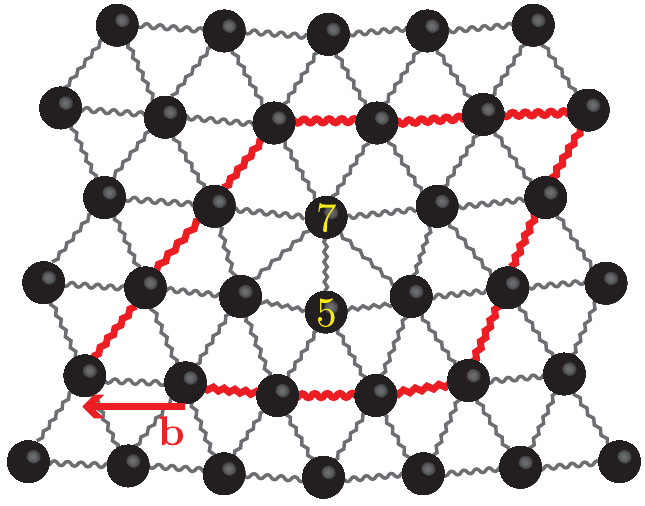}
\caption{Dislocation in a triangular lattice. The Burger's vector
  specifies by how much a clockwise circuit (marked in red, bold) around the dislocation
  fails to close. A dislocation can be viewed as disclination dipole
  with a moment perpendicular to its Burger's vector.}
\label{fig:dislocation}
\end{figure}

Besides disclinations, dislocations can occur in crystals. Dislocations are characterised by a Burger's vector $\mathbf{b}$. This vector measures the change in the displacement vector, if we make a counterclockwise loop surrounding the dislocation,  
\begin{equation}
\label{eq:dislocation}
\oint d \mathbf{u} = \mathbf{b}. 
\end{equation}
Just like the strength of disclinations can only take on a value out
of a discrete set, the Burger's vector of a dislocation is equal to
some integer multiple of a lattice vector. Also note that a dislocation can be viewed as a pair of closely spaced disclinations of opposite charge \cite{PhysRevB.19.2457}, as can be seen in Fig. \ref{fig:dislocation}. 

The flat space differential form of eq. (\ref{eq:dislocation}) for a
dislocation at $\mathbf{x}_a$ is
\begin{equation}
\label{eq:dislocation_diff}
\epsilon_{ij} \partial_i \partial_j u_k = b_k \delta \left( \mathbf{x} - \mathbf{x}_a\right), 
\end{equation}
which again can be obtained by using Stokes' theorem.


\section{Interaction between curvature and defects}
\subsection{Coupling in liquid crystals}
It is possible to recast the free energy in terms of the locations of
the topological defects rather than the director or displacement
field, if smooth (\textit{i.e.} non-singular) deformations are
ignored. In this case the energy in
eq. \eqref{eq:energy with connection} is minimised with respect to $\Theta$, leads to 
\begin{equation}
\label{eq:lc_minimisation}
 D^\alpha \left( \partial_\alpha \Theta - A_\alpha \right) = 0.
 \end{equation}
This needs to supplemented with an equation for the effective charge distribution: 
\begin{equation}
\label{eq:rhominG}
\gamma^{\alpha \beta} D_\alpha \left( \partial_\beta \Theta - A_\beta \right) = \rho - G,
\end{equation}
obtained by combining eq. \eqref{eq:curl_connection_gauss}
for the curvature and eq. \eqref{eq:disclination_nematic_diff} for the defect density $\rho \left( \mathbf{x} \right)$, 
\begin{equation}
\rho = \sum_a q_a \delta \left( \mathbf{x} - \mathbf{x}_a \right).
\end{equation}
Eq. \ref{eq:lc_minimisation} is automatically satisfied if one chooses \cite{PhysRevE.70.051105} 
\begin{equation}
\label{eq:lc_chi}
\partial_\alpha \Theta - A_\alpha = \gamma_\alpha^{\phantom{\alpha}\beta} \partial_\beta \chi,
\end{equation} 
where $\chi \left( \mathbf{x} \right)$ is an auxiliary function. 
At the same time, substituting eq. \eqref{eq:lc_chi} into
eq. \eqref{eq:rhominG} leads to
\begin{equation}
\label{eq:lc_poisson}
-D^2 \chi = \rho - G.
\end{equation}
The source in this Poisson equation contains both topological point
charges as well as the Gaussian curvature with opposite sign. The
analog of the electrostatic potential is $\chi$. The role of the
electric field is played by $\partial_\alpha \chi$. Indeed,
substituting eq. \eqref{eq:lc_chi} in eq. \eqref{eq:energy with connection},
shows that the energy density is proportional to the square of the
electric field:
\begin{equation}
\label{eq:lc_Fchi}
F =\frac{1}{2} k \int dS  \partial_{\alpha}
 \chi \partial^{\alpha} \chi. 
\end{equation}
Next, we formally
solve eq. \eqref{eq:lc_poisson} 
\begin{equation}
\label{eq:lc_chi_sol}
\chi = -\int dS' \Gamma_{L} \left( \mathbf{x}, \mathbf{x'} \right) \left( \rho \left( \mathbf{x'} \right) - G \left( \mathbf{x'} \right) \right)
\end{equation}
where $\Gamma_{L} \left( \mathbf{x}, \mathbf{x'} \right)$ is the Green
function of the Laplace-Beltrami operator, $D^2 = D_\alpha D^\alpha$, satisfying
\begin{equation}
D^2 \Gamma_{L} \left( \mathbf{x}, \mathbf{x'} \right) =  \delta \left( \mathbf{x} - \mathbf{x'}\right). 
\end{equation}
Integrating eq.\eqref{eq:lc_Fchi} by parts and substituting our
expressions for $\chi$ and the Laplacian of $\chi$
(eqs. \eqref{eq:lc_chi_sol} and \eqref{eq:lc_poisson} respectively)
results (up to boundary terms) in 
\begin{equation}
\label{eq:energy_LC_coupling}
F = -\frac{k}{2} \int dS \int dS' \left( \rho \left( \mathbf{x} \right) - G \left( \mathbf{x} \right) \right) \Gamma_{L}\left( \mathbf{x}, \mathbf{x'} \right) \left( \rho \left( \mathbf{x'} \right) - G \left( \mathbf{x'} \right) \right) ,
\end{equation}
from which we again deduce the analogy with two-dimensional
electrostatics. In this analogy the defects are electric point sources
with their electric charge equal to the topological charge $q$ and the
Gaussian curvature with its sign reversed is a background charge
distribution. Therefore the defects will be attracted towards regions
of Gaussian curvature with the same sign as the topological charge
\cite{PhysRevLett.67.1169, PhysRevE.69.041102,PhysRevE.70.051105, PhysRevLett.93.215301,
  Xing03042012, Jesenek2012277,doi:10.1021/jp205128g,0253-6102-46-2-028}. Such screening will be perfect if $S=\rho$
everywhere, since $F=0$ then. However, unless the surface contains
singularities in the Gaussian curvature, like the apex of a cone,
perfect screening
will be impossible, as the topological charge is quantised whereas the Gaussian
curvature is typically smoothly distributed.

\subsection{Coupling in crystals}
\begin{figure}[h]
\centering
\subfloat{\label{fig:hex5}\includegraphics[width=0.4\textwidth]{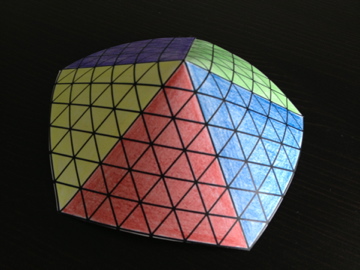}}                
\subfloat{\label{fig:hex7}\includegraphics[width=0.4\textwidth]{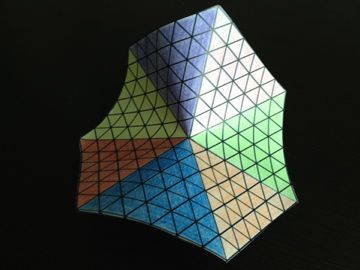}}\\
\vspace{-0.37cm}
\subfloat{\label{fig:sq3}\includegraphics[width=0.4\textwidth]{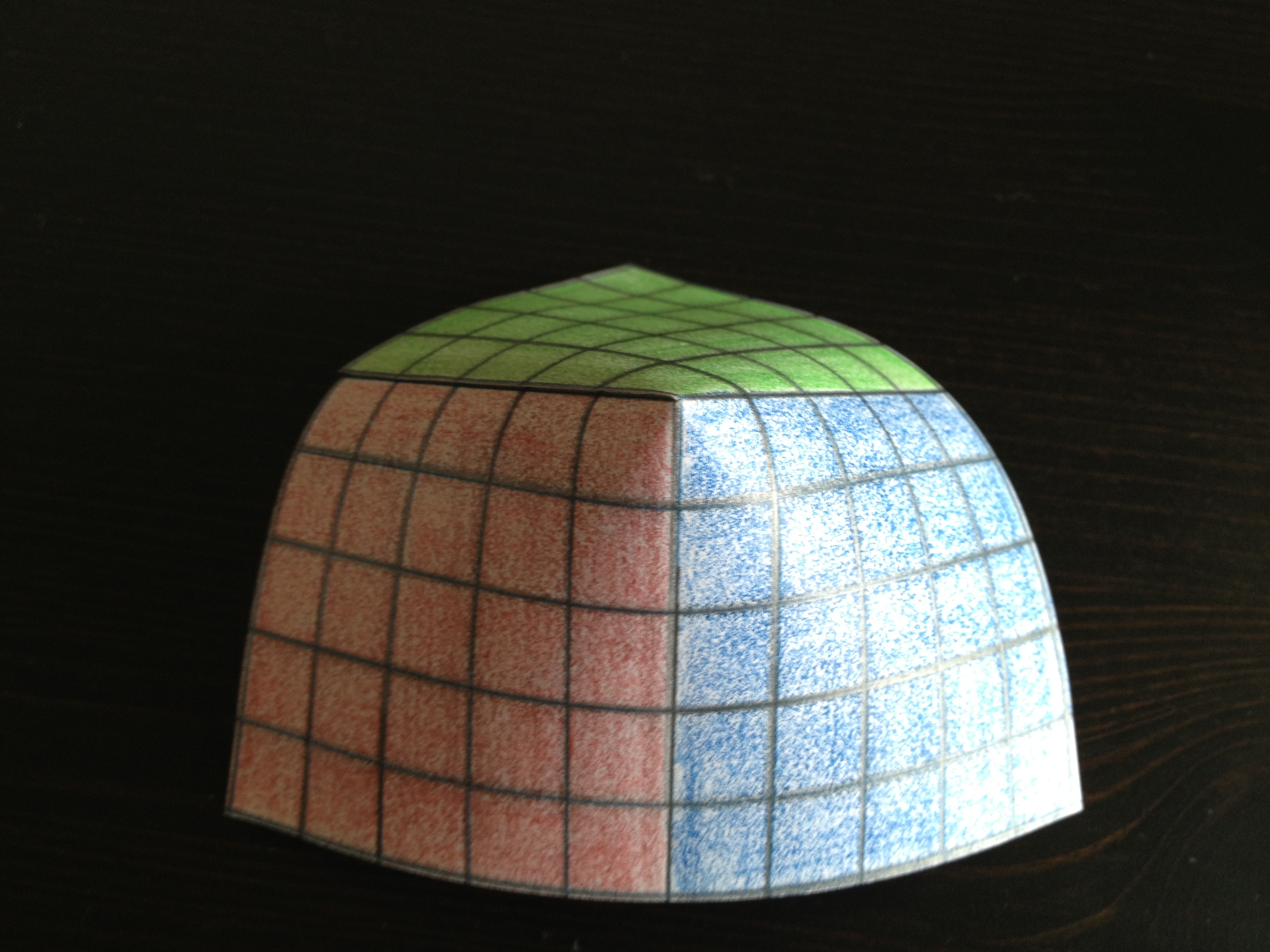}}                
\subfloat{\label{fig:sq5}\includegraphics[width=0.4\textwidth]{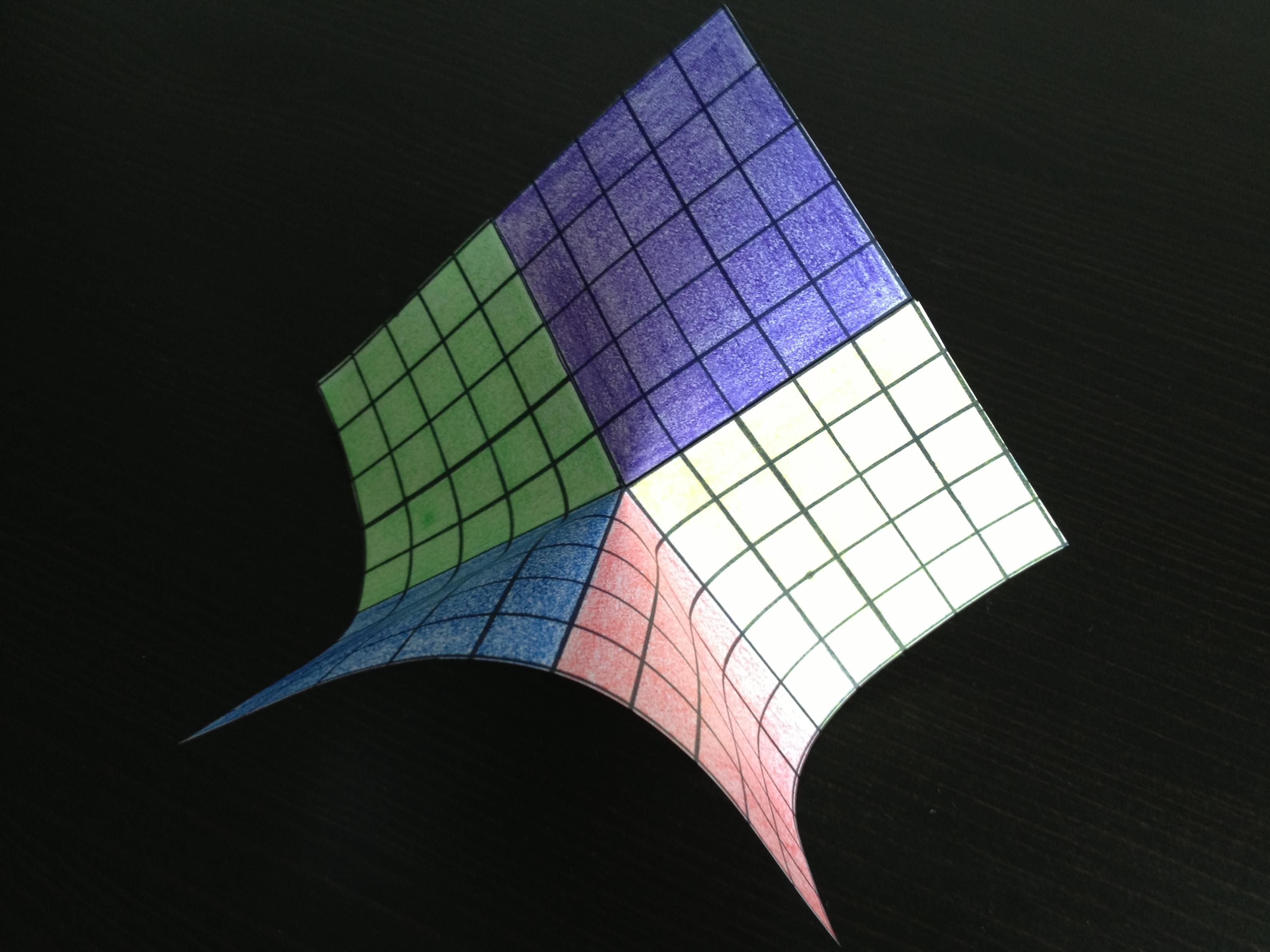}}
\caption{Paper models illustrating the coupling between disclinations
  and curvature. \textit{Left panels:} Positively (\textit{right panels}: negatively)
  charged disclinations and positive (negative) Gaussian curvature
  attract. \textit{Top left panel:} 5-fold coordinated particle in a triangular lattice. \textit{Top right
  panel:} 7-fold coordinated particle in a triangular lattice. \textit{Bottom
  left panel:} 3-fold coordinated particle in a square
lattice. \textit{Bottom right panel:}
  5-fold coordinated particle in a square lattice.
\label{fig:curvature_disclination}}
\end{figure}

Note that an \textit{arbitrary} field $\chi$ solves eq.
\eqref{eq:force_balance_solved}. However, $\chi$ must be physically
possible and we therefore need to accompany eq.
(\eqref{eq:force_balance_solved}) with another equation, which we will
obtain by considering the inversion of eq. \eqref{eq:Hooke} \cite{Landau,PhysRevA.38.1005}:
\begin{align}
u_{ij} &= \frac{1+\nu}{Y}\sigma_{ij} - \frac{\nu}{Y} \sigma_{kk} \delta_{ij} \\
&= \frac{1+\nu}{Y} \epsilon_{ik} \epsilon_{jl} \partial_k \partial_l
\chi - \frac{\nu}{Y} \partial^2 \chi \delta_{ij} \label{eq:strain_and_chi}
\end{align}
where the two-dimensional Young's modulus, $Y$, and Poisson ratio, $\nu$, are given by
\begin{align}
Y &= \frac{4 \mu \left( \mu + \lambda \right) }{2\mu + \lambda}, \\
\nu &= \frac{\lambda}{2\mu + \lambda}.
\end{align}
Applying $ \epsilon_{ik} \epsilon_{jl} \partial_k \partial_l $ to
eq.  \eqref{eq:strain_and_chi} gives 
\begin{equation}
\frac{1}{Y} \partial^4 \chi = \epsilon_{ik} \epsilon_{jl} \partial_k \partial_l u_{ij}.
\end{equation}
By invoking eqs. \eqref{eq:strain}, \eqref{eq:bondangle}, the
differential expressions for the defects, namely eqs.
\eqref{eq:dislocation_diff} and \eqref{eq:disclination_crystal_diff},
as well as eq. \eqref{eq:A_and_curvature} for the curvature,
one can rewrite the right hand side to arrive at the crystalline
analog of eq. \eqref{eq:lc_poisson}:
\begin{equation}
\label{eq:source_airy}
\frac{1}{Y} \partial^4 \chi = \rho - G,
\end{equation}
where the defect distribution, $\rho$, of disclinations with charge $q_a$ and dislocations with Burger's vector $\mathbf{b}^b$ reads
\begin{equation}
\rho = \sum_a q_a \delta \left( \mathbf{x} - \mathbf{x}_a \right)  + \sum_{b} \epsilon_{ij} b_{i}^{b} \partial_j \delta \left( \mathbf{x} - \mathbf{x}_b \right).
\end{equation}
We can also rewrite the free energy (up to boundary terms) in terms of the Airy stress function as follows:
\begin{equation}
F = \frac{1}{2Y} \int dS \left( \partial^2 \chi \right)^2
\end{equation}
If we integrate this by parts twice and use eq.
\eqref{eq:source_airy} to eliminate $\chi$ and $\partial^4 \chi$, we find (up to boundary terms)
\begin{equation}
\label{eq:energy_crystal_coupling}
F = \frac{Y}{2} \int dS \int dS' \left( \rho \left( \mathbf{x} \right) - G \left( \mathbf{x} \right) \right) \Gamma_{B} \left( \mathbf{x}, \mathbf{x'} \right) \left( \rho \left( \mathbf{x'} \right) - G \left( \mathbf{x'} \right) \right)  
\end{equation}
where $\Gamma_{B}$ is the Greens function of the biharmonic operator
\begin{equation}
\partial^4 \Gamma_{B} \left( \mathbf{x}, \mathbf{x'} \right) =  \delta \left( \mathbf{x} - \mathbf{x'}\right). 
\end{equation}
Eq. \eqref{eq:energy_crystal_coupling} is the crystalline analog
of eq. \eqref{eq:energy_LC_coupling}. Again, the defects can
screen the Gaussian curvature. The interaction, however, is different
than the Coulomb interaction in the liquid crystalline case. If the
surface is allowed to bend, disclinations will induce buckling,
illustrated in Fig. \ref{fig:curvature_disclination} with paper models.
In these cones, the integrated Gaussian curvature is determined by the angular
deficit of the disclination
\begin{equation}
\int dS G = q.
\end{equation}

\subsection{Screening by dislocations and pleats}
Surprisingly, also charge neutral dislocations and pleats can screen
the curvature \cite{refId0,PhysRevA.38.1005,Vitelli15082006,PhysRevE.76.051604,2010Natur.468..947I}. Pleats are formed by arrays of dislocations and allow
for an extra piece of crystal, just like their fabric analogs. The
opening angle, $\Delta \Theta$, of the pleat (or low angle grain boundary) is given by
\begin{equation}
\Delta \Theta \approx n_d a 
\end{equation}
where $a$ is the lattice spacing and $n_d$ is the dislocation line density. 
Since this opening angle can be arbitrarily small, pleats can provide a finer screening than quantised disclinations. 

\subsection{Geometrical potentials and forces}
The cross terms of equation (\ref{eq:energy_crystal_coupling}) represent the interaction energy
\begin{equation}
\label{eq:zeta}
\zeta = - Y \int dS \rho \left( \mathbf{x} \right) \int dA'
\Gamma_{B}\left( \mathbf{x}, \mathbf{x'} \right)  G \left( \mathbf{x'}
\right) 
\end{equation}
By introducing an auxiliary function $V \left( \mathbf{x} \right)$
satisfying
\begin{equation}
 \partial^2 V =  G,
\end{equation}
eq. \eqref{eq:zeta} can by integrating by parts twice be rewritten as
\begin{equation}
\zeta = - Y \int dS \rho \left( \mathbf{x} \right) \int dS'
\Gamma_{L}\left( \mathbf{x}, \mathbf{x'} \right)  V \left( \mathbf{x'}
\right) 
\end{equation}
The field $\zeta \left( \mathbf{x} \right)$ can be viewed as a
geometric potential, \textit{i.e.} the potential experienced
by a defect due to the curvature of the crystal \cite{Vitelli15082006,2009AdPhy..58..449B}. Another, more
heuristic way, to study the interaction of dislocations and curvature
is the following. We consider the stress that exist in the monolayer
as a result of curvature only, $\sigma_{ij}^{G}$, as the source of a Peach-Koehler force, $\mathbf{f}$, on the dislocation:    
\begin{equation}
f_k = \epsilon_{kj} b_i \sigma_{ij}^{G}.
\end{equation}
Note that, by setting $\rho=0$, the Airy stress function $\chi^{G}$ satisfies
\begin{equation}
\frac{1}{Y} \partial^4 \chi^G =  - G
\end{equation}
This equation can be solved in two steps. First, we make use of an auxiliary function $U$ obeying
\begin{equation}
\partial^2 U = G
\end{equation}
This leaves the following equation to be solved
\begin{equation}
\frac{1}{Y} \partial^2 \chi^G = - U + U_H, 
\end{equation}
where $U_H$ is a harmonic function (\textit{i.e.} $\partial^2 U_H =
0$) introduced to fulfil the boundary conditions \cite{Vitelli15082006}.

\section{Nematics in spherical geometries}
\subsection{Nematic order on the sphere}
\begin{figure}[h]
\centering
\includegraphics[width=0.4\textwidth]{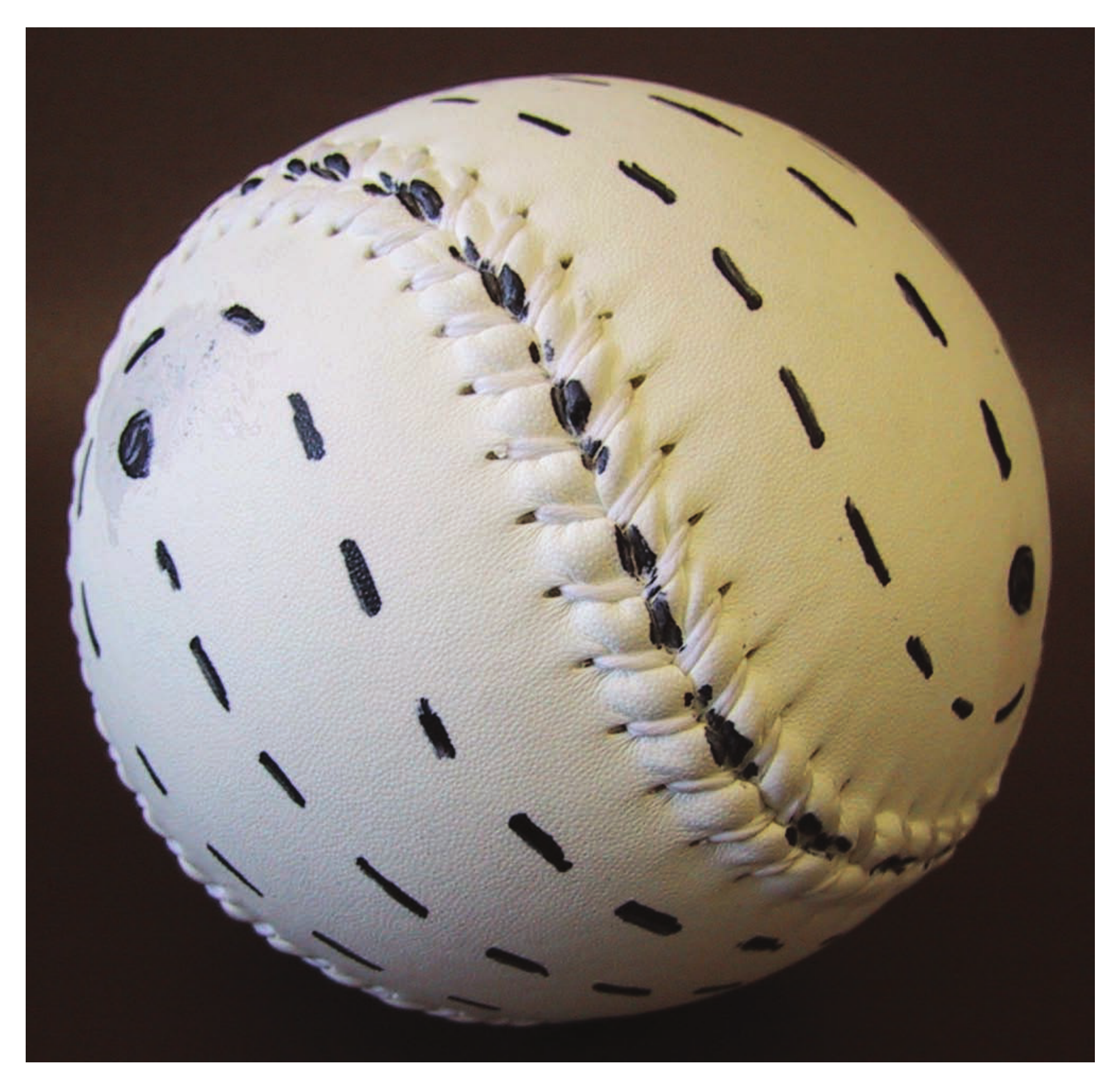} 
\caption{The baseball-like ground state of a two-dimensional spherical
  nematic coating has four $s=\frac{1}{2}$ at the vertices of a
  tetrahedron in the one-constant approximation. Figure from
  \cite{2006PhRvE..74b1711V}. \label{fig:baseball}}
\end{figure}

As a naive guess for the ground state of a two dimensional nematic
liquid crystal phase on the surface of the sphere, one could imagine
the excess of topological charge to be located at the poles, like in
the case of tilted molecules on the sphere. However, the order
parameter, the director, has the symmetry of a headless arrow instead
of a vector. Therefore, this makes it possible for the two $s=1$
defects to unbind into four $s=\frac{1}{2}$ defects relaxing at the
vertices of a regular tetrahedron \cite{1992JPhy2...2..371L}. The
baseball-like nematic texture is illustrated in
Fig. \ref{fig:baseball}. 
The repulsive nature of defects with like charges can be seen from the free energy, which, as shown in the previous section, can entirely be reformulated in terms of the defects rather than the director \cite{2002NanoL...2.1125N, 1992JPhy2...2..371L}:
\begin{equation}
F = -\frac{\pi k}{2}\sum_{i \neq j} s_i s_j \log \left( 1 - \cos \theta_{ij} \right) + E\left(R\right) \sum_j s_j^2.
\end{equation}
Here, $\theta_{ij}$ is the angular separation between defects $i$ and $j$, i.e. $\theta_{ij}=\frac{d_{ij}}{R}$, with $d_{ij}$ being the geodesic distance. The first term yields the long-range interaction of the charges. The second term accounts for the defect self-energy
\begin{equation}
E\left(R\right) = \pi k \log \left( \frac{R}{b} \right)+E_c,
\end{equation}
where we have imposed a cut-off $b$ representing the defect core size,
which has energy $E_c$. This cut-off needs to be
introduced in order to prevent the free energy from
diverging. Heuristically, this
logarithmically diverging term in the free energy is responsible for
the splitting of the two $s=1$ defects into four $s=\frac{1}{2}$
defects. Two $s=1$ defects contribute $\left( 2 \times 1^2\right) \pi
k \log \left( \frac{R}{b} \right)= 2\pi k \log \left( \frac{R}{b}
\right)$ to the free energy, whereas four $s=\frac{1}{2}$ defects
contribute only $\left(4 \times \left(\frac{1}{2}\right)^2 \right) \pi k \log \left( \frac{R}{b} \right)=\pi k \log \left( \frac{R}{b} \right)$.

In addition to this ground state, other defect structures have been observed in computer simulations \cite{PhysRevE.62.5081, PhysRevLett.100.197802, 2008PhRvL.101c7802S, 2008JChPh.128j4707B}. If there is a strong anisotropy in the elastic moduli, the four defects are found to lie on a great circle rather than the vertices of a regular tetrahedron \cite{2008PhRvL.101c7802S, 2008JChPh.128j4707B}.

\subsection{Beyond two dimensions: spherical nematic shells}
An experimental model system of spherical nematics are nematic double
emulsion droplets \cite{2007PhRvL..99o7801F, PhysRevE.79.021707,
  2011NatPh...7..391L, 2011PhRvL.106x7801L, 2011PhRvL.106x7802L, 2012PhRvE..86b0705S,
  C2SM07415J, C3SM27671F, Liang13042013}. These are structures in which a water droplet is
captured by a larger nematic liquid crystal droplet, which in turn is
dispersed in an outer fluid. There are some crucial differences
between a two-dimensional spherical nematic and these systems. Not
only is the nematic coating of a finite thickness, this thickness can
be inhomogeneous as a result of buoyancy driven
displacement (or other mechanisms) of the inner droplet out of
the centre of the nematic droplet. 

\begin{figure}[h]
\subfloat
{\includegraphics[width=0.4\columnwidth]{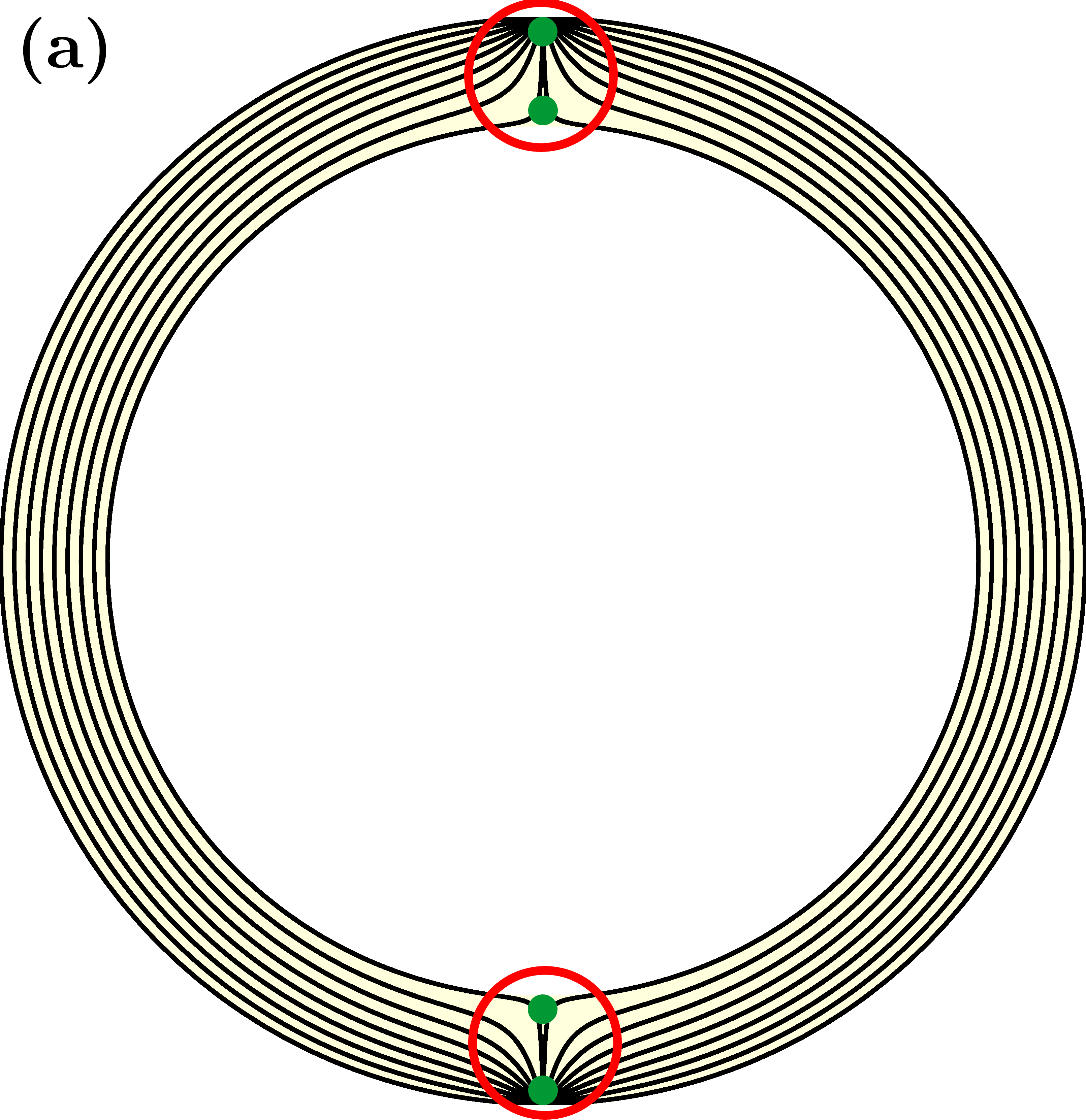}
\label{fig:introa}}
\subfloat 
{\includegraphics[width=0.4\columnwidth]{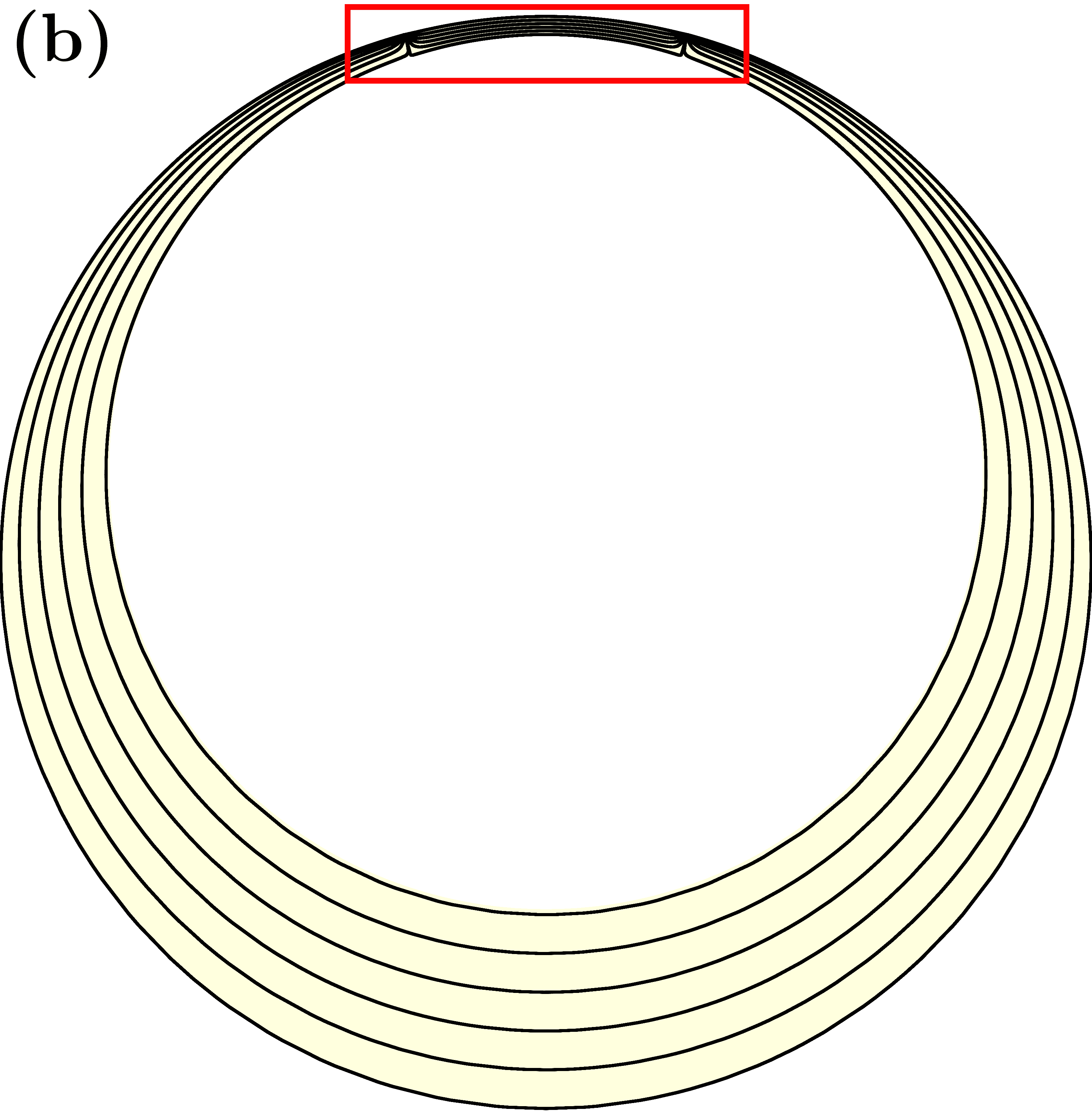}
\label{fig:introb}}\\
\subfloat 
{\includegraphics[width=0.8\columnwidth]{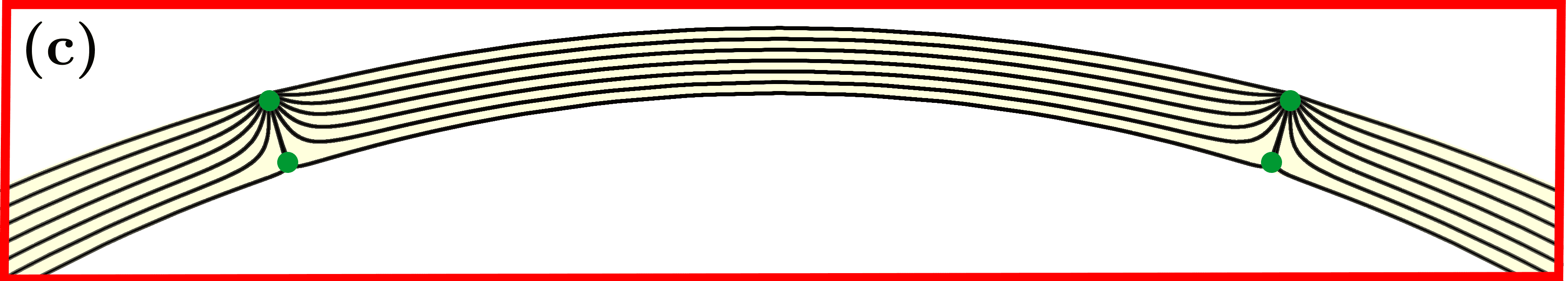}
\label{fig:introc}}
\caption{(a) Schematic of the deconfined defect configuration in
  a homogeneous shell. Two pairs (each encircled in red) of boojums,
  indicated by green dots, are located at the top and bottom of the
  shell. (b) Schematic of the confined defect configuration in
  an inhomogeneous shell. All boojums are located at the thinnest, top
  part of the shell, inside the red rectangle.  (c) Zoom of the thinnest section of the
  inhomogeneous shell in (b). From \cite{C3SM27671F} - Reproduced by permission of The Royal Society of Chemistry.\label{fig:boojum_in_shell_confined_and_deconfined}}
\end{figure}

Like point disclinations in two dimensions, there exist disclination \textit{lines} in a three dimensional nematic liquid crystal, which are categorised in similar fashion. 
However, charge one lines, and integral lines in general, do not
exist. Such lines loose their singular cores \cite{Cladis,
  1973PMag...27..405M} by `escaping in
the third dimension'. In shells, such an escape leads to another type
of defects, namely point defects at the interface, known as
boojums (Fig. \ref{fig:boojum_in_shell_confined_and_deconfined}). 

In a spherical nematic layer of \textit{finite}
thickness, calculations show that the baseball structure with four
$s=\frac{1}{2}$ disclination lines spanning the shell, become
energetically less favourable than two antipodal pairs of boojums
beyond a critical thickness \cite{2006PhRvE..74b1711V}. Instead of
unbinding, the singular lines escape in the third dimension,
leaving two pairs of boojums on the bounding surfaces. These two
defect configurations are separated by a large energy barrier. As a
consequence, both configurations are observed in droplets in the same emulsion.  If, in
addition, the shell thickness is inhomogeneous, the energy landscape
becomes even more complex.

As a consequence of the inhomogeneity the defects cluster in the
thinnest part of the shell, where the length of the disclination lines
(or distance between boojums forming a pair) are shorter. Since the self-energy of the disclination is proportional to its length, it is attracted towards this region of the shell.     
One of the intriguing outcomes of the study of inhomogeneous shells is
that in the two defects shell, the pairs of surface defects can make abrupt
transitions between the state in which the defects are confined in the
thinnest part of the shell, and the deconfined state, in which the
interdefect repulsion places them diametrically \cite{C3SM27671F}. These confinement and
deconfinement transitions occur when the thickness or thickness
inhomogeneity is varied. A defect arrangement with a corresponding local
minimum in the energy landscape makes the transition to the global
minimum when the local minimum looses its stability. This explains
both the abruptness of the transitions as well as the hysteresis
between them.     

In agreement with this picture, Monte Carlo simulations of nematic shells on uniaxial and
biaxial colloidal particles have shown the tendencies for defects to
accumulate in the thinnest part and in regions of the highest
curvature \cite{B917180K}.

\section{Toroidal nematics}
\begin{figure}[h]
\centering
\includegraphics[width=0.7\textwidth]{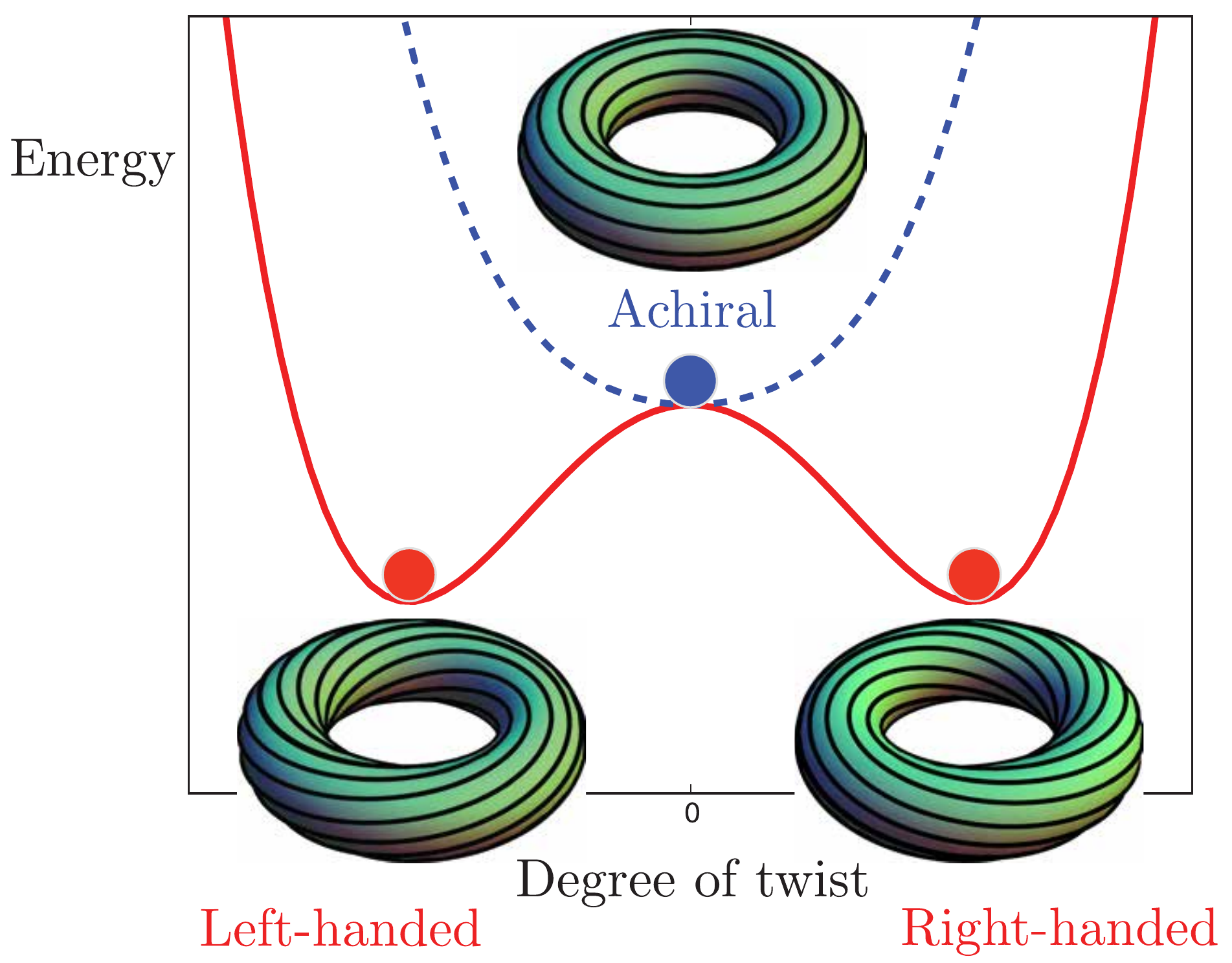} 
\caption{Energy as a function of the degree of twist has either a
  single achiral minimum (dashed blue) or shows spontaneous
  chiral symmetry breaking in toroidal nematics (red) depending on
  the aspect ratio and elastic constants. The chiral state is
  favoured for fat toroids and small values of $\left(K_2-K_{24}\right)/K_3$. \label{fig:chiral_symmetry_breaking}}
\end{figure}

The torus has a zero topological charge. Hence, in a nematic droplet of toroidal
shape no defects need to be present. The director
field to be expected naively in such a geometry is one which follows the
tubular axis, as shown in Fig. \ref{fig:chiral_symmetry_breaking}. This achiral director configuration contains
only bend energy. Simple analytical calculations show, however, that
if the toroid becomes too fat it is favourable to reduce
bend deformations by
twisting. The price of twisting is screened by saddle-splay
deformations provided that $K_{24}>0$ \cite{Pairam04062013,2013arXiv1312.5092K}. The twisted configuration is chiral. Chirality stems from the
 Greek word for hand, and is indeed in this context easily explained:
 your right hand cannot be turned into a left hand by moving and
 rotating it. It is only when viewed in the mirror that your right hand
 appears to be a left hand and vica versa. Indeed, for small aspect ratios and small values of $\left(K_2-K_{24}\right)/K_3$ nematic toroids
 display either a right- or left-handedness despite the achiral
 nature of nematics. This phenomenon is recognised as spontaneous chiral
 symmetry breaking. Typical corresponding plots of the energy as a function of the
 amount of twist are shown in  Fig. \ref{fig:chiral_symmetry_breaking}. 

\section{Concluding remarks}
We hope to have shared our interest in the rich subject of geometry in
soft matter, in particular the interplay of defects and curvature in
two-dimensional ordered matter and the confinement of nematic liquid
crystals in various geometries. For readers interested in a more
detailed treatment, we refer to
excellent reviews by Kamien\cite{RevModPhys.74.953}, Bowick and
Giomi\cite{2009AdPhy..58..449B}, Nelson\cite{Defects_and_Geometry},
David\cite{David}, and Lopez-Leon and Fernandez-Nieves \cite{Lopez-Leon:2011fk}.


\bibliography{liquid_crystals} 
\end{document}